\documentclass[useAMS, twocolumn, aps, prd, amscd, amsmath, amssymb,verbatim,nofootinbib,usenatbib]{revtex4}
\usepackage{times}
\usepackage{graphicx}
\usepackage{verbatim}
\usepackage{tikz}
\usepackage{color}

\usepackage[T1]{fontenc}
\usepackage{aecompl}
\usepackage{calligra}
\DeclareMathAlphabet{\mathcalligra}{T1}{calligra}{m}{n}
\DeclareFontShape{T1}{calligra}{m}{n}{<->s*[2.2]callig15}{}

\usepackage{hyperref}
\hypersetup{
    pdfnewwindow=true,      % links in new window
    colorlinks=true,       % false: boxed links; true: colored links
    linkcolor=blue,          % color of internal links
    citecolor=blue,        % color of links to bibliography
    filecolor=blue,      % color of file links
    urlcolor=blue           % color of external links
}

\newcommand{\scripty}[1]{\ensuremath{\mathcalligra{#1}}}

\def\Res{\mathcal{R}}

\def\Evec{{\bf E}}
\def\Bvec{{\bf B}}

\def\orb{\rm{orb}}
\def\Msun{{M_{\odot}}}

\def\MBH{M}

\def\Lum{\mathcal{P}}

\def\rns{r}
\def\rbh{\scripty{r}}
\def\Rbin{r}
\def\RNS{R_{\rm NS}}

\def\bea{\begin{eqnarray}}
\def\eea{\end{eqnarray}}

\def\be{\begin{equation}}
\def\ee{\end{equation}}

\begin{document}
\title[Bright transients from neutron star-black hole mergers]{Bright transients from strongly-magnetized neutron star-black hole mergers}
\author{Daniel J. D'Orazio}
\email{dorazio@astro.columbia.edu}
\affiliation{Department of Astronomy, Columbia University, New York, New York 10027, USA}

\author{Janna Levin}
\affiliation{Department of Physics and Astronomy,
Barnard College, Columbia University, New York, New York 10027, USA}

\author{Norman W. Murray}
\thanks{Canada Research Chair in Astrophysics}
\affiliation{Canadian Institute for Theoretical Astrophysics, 60
Street George Street, University of Toronto, Toronto, Ontario M5S 3H8, Canada}

\author{Larry Price}
\affiliation{Division of Physics, Mathematics, and Astronomy, California Institute of Technology, Pasadena, California 91125, USA}

\begin{abstract}
Direct detection of black hole-neutron star pairs is
anticipated with the advent of aLIGO.  Electromagnetic counterparts
may be crucial for a confident gravitational-wave detection as well as
for extraction of astronomical information. Yet black hole-neutron star pairs are
notoriously dark and so inaccessible to telescopes.  Contrary to this
expectation, a bright electromagnetic transient can occur in the final
moments before merger as long as the neutron star is highly
magnetized.  The orbital motion of the neutron star magnet creates a
Faraday flux and corresponding power available for luminosity. A
spectrum of curvature radiation ramps up until the rapid injection of
energy ignites a fireball, which would appear as an energetic
blackbody peaking in the x ray to $\gamma$ rays for neutron star
field strengths ranging from $10^{12}$G to $10^{16}$G respectively and
a $10M_{\odot} $ black hole. The fireball event may last from a few
milliseconds to a few seconds depending on the neutron star magnetic-field
strength, and may be observable with the Fermi Gamma-Ray Burst Monitor
with a rate up to a few per year for neutron star field strengths
$\gtrsim 10^{14}$G.  We also discuss a possible decaying post-merger
event which could accompany this signal.  As an
electromagnetic counterpart to these otherwise dark pairs, the black-hole battery
should be of great value to the development of multi-messenger
astronomy in the era of aLIGO.
\end{abstract}

\maketitle

\section{introduction}
Black holes are dark dead stars. Neutron stars are giant magnets. As
the neutron star (NS) whips around the black hole (BH) in the final
stages in the life of a pair, an electromotive force (emf) is
generated that is powerful enough to light a beacon, which conceivably we
might observe at cosmological distances
\citep{McL:2011,DorazioLevin:2013}. The battery could power
synchrocurvature radiation, a blazing fireball, or relativistic jets.

Famously, tidal disruption of a NS is expected to generate a gamma-ray
burst after merger \citep{NPP:NSBH_GRB:1992}. However, it is
under-appreciated that most BHs should be large enough ($\gtrsim
6M_\odot $) to swallow their NSs whole and so no gamma-ray burst is
expected from typical pairs \citep{Ozel:2010}. Therefore, our BH
battery, which operates with the NS intact, may be one of the only
significant sources of electromagnetic luminosity for coalescing BHNS binaries.\footnote{Resonant shattering of the NS crust could also generate an interesting electromagnetic 
signature for nondisrupting systems \citep{Tsang:2012, Tsang:2013}.}
An observation of such a transient would be exciting in its own right.
Advanced gravitational-wave detectors \citep[{\em
    e.g.},][]{AdLIGO:2010}, with the prospect of multi-messenger
astronomy, provide added incentive for the more detailed predictions
of the electromagnetic (EM) signatures we present here.

Even with the benefit of nearly fifty years of observations, common NS
pulsars require theoretical attention. If the decades of pulsar
research offer a sociological lesson, it would be that the details of
the electromagnetic processes are not easy to model, that the
mechanisms at work are not obvious. Without the benefit of
observations, we would not presume to offer a definitive or complete
electromagnetic portrait of the BHNS engine. But we can sketch
plausible emission mechanisms to encourage first searches for these
potentially important transients.

As already argued in the original references
\citep{McL:2011,DorazioLevin:2013}, curvature radiation is a natural
channel for luminosity.  We examine the spectrum of curvature
radiation here.  (We mention that another intriguing channel for some
fraction of the battery power could be radio emission through coherent
processes, providing the correct time scales and energetics for a
subclass of the fast radio bursts \citep{Chiara:2015}.)  We conclude
that, just before merger, when the power is greatest, curvature
radiation results in copious pair production which fuels a
fireball. The fireball expands under its own pressure until the
photosphere radiates as a blackbody peaking in the hard x-ray to
$\gamma$-ray range for milliseconds (msec) to seconds depending on NS
magnetic-field strength.  

If the merger were to happen in our own
galaxy, we might watch the spectrum of curvature radiation ramp up
followed by the brighter fireball. At cosmological distances, the
high-energy lead up in curvature radiation will be too faint to
detect, but the fireball could be observable at a rate of at least a
few per year with the FERMI Gamma-Ray Burst Monitor (GBM), for NSs
with $\gtrsim10^{14}$G surface magnetic fields.  Such events 
could possibly be a subclass of short gamma-ray
bursts. Since the fireball takes at least $\sim 0.2$ms to $0.02$s 
to expand and release the light, the burst from the
fireball would lag just behind the peak gravitational-wave emission.
Post-merger, the transfer of magnetic flux on to the black hole might
lead to a brief jet and afterglow. 
Pre- and post-merger triggered events could be observed to occur very close to each other in timing.
We hope the predicted transient discussed here encourages observational interest.

\subsection{The power of the battery}
\label{Orbital inspiral battery}
First, we review the estimate of the energy budget for the BH
battery. The BHNS system behaves analogously to a unipolar inductor, which has been investigated in application 
to a number of other astrophysical systems, \textit{e.g.} Jupiter and 
its moon Io \citep{GLB:1969}, planets around white dwarfs \citep{Li:1998} and main sequence stars \citep{LaineLinI:2012,LaineLinII:2012}, 
binary neutron stars \citep{Vietri:1996,Piro:2012, DLai:2012, Palenzuela:2013}, 
compact white dwarf binaries \citep{Wu:2002, Dall'Osso:2006, Dall'Osso:2007, 
DLai:2012}, BHs boosted through magnetic fields 
\citep{Lyut:2011, Penna:2015}, and the Blandford-Znajek (BZ) mechanism \citep{BZ:1977} for a single BH spinning in a magnetic field \citep[for recent numerical work on the BZ mechanism see \textit{e.g.}][]{PalenzuelaBZ:2011, Kiuchi:2015}. The calculation for BHNS systems, 
already presented in Ref.\ \cite{McL:2011}
and confirmed in the detailed relativistic analysis of
Ref.\ \cite{DorazioLevin:2013}, as well as the numerical calculations
of Ref.\ \cite{Paschalidis:2013}, gives the scaling of power available
for conversion into electromagnetic luminosity. In the next section we
will consider the implications of throwing this power into luminous
elements in the BHNS circuit.

\begin{figure}
\begin{center}
\includegraphics[scale=0.33]{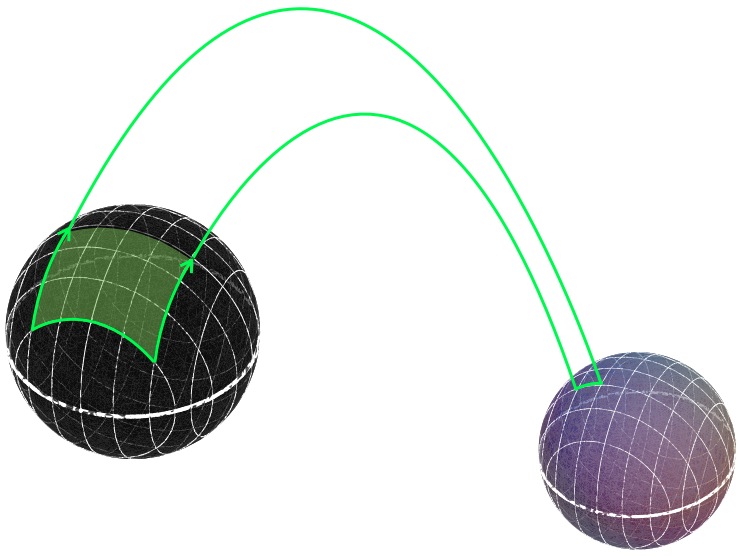} 
\end{center}
\caption{Schematic of a Faraday loop as seen by an observer external
  to the horizon. The black sphere depicts the BH horizon orbiting out
  of the page. In green is a schematic of the instantaneous closed
  loop defining one of infinitely many circuits made up of electrons
  and positrons moving along magnetic-field lines which trace the BH
  horizon.}
\label{Fig:Contour}
\end{figure}

For observers which have not fallen through, the BH horizon is well
approximated, electromagnetically, as a conducting sphere
\citep{MPBook}. The relative motion of the BH through the magnetic
field of the NS induces an emf. We visualize the circuit which
generates this emf in Figure\ \ref{Fig:Contour}. Because charged
particles are bound to a given field line, we imagine that one set of
field lines forms one set of wires in a closed circuit. In
conceptualizing the circuit it is important to distinguish between
field lines that act as wires at a given instant and those that
contribute to the changing magnetic flux through the circuit. The
circuit is closed by connecting the wires along the surface of the
horizon, as in the snapshot of Figure\ \ref{Fig:Contour}.  As the BHNS
pair orbits, the circuit sweeps through the dipole field. The changing
magnetic flux through a surface bounded by the changing circuit
corresponds to an emf. There are an infinite number of such circuits
as different field lines intersect the BH.

%%%%%%%%%%%%%%%%%%%%%%%%%%%%%%%%%%
%%%FIGURE Faraday Loop
%%%%%%%%%%%%%%%%%%%%%%%%%%%%%%%%%%
%%%%%%%%%%%%%%%%%%%%%%%%%%%%%%%%%%%%%%%%%%%%%%%%

Following Ref.\ \cite{McL:2011}, the voltage generated is given by
\begin{align}
V_{\mathcal{H}} = \int{ \alpha \mathbf{E} \cdot \mathbf{ds} } &= 
- \frac{1}{c} \frac{d}{dt} \int{ \alpha   \mathbf{B} \cdot \mathbf{dA} } \nonumber \\
&= -\oint{ \alpha \left( \frac{\mathbf{v}}{c} \times \mathbf{B} \right) \cdot \mathbf{ds} },
\label{Eq:Volts}
\end{align}
where $\mathbf{v}$ is the relative velocity of the BH horizon with respect to magnetic-field lines and we add a factor of the lapse function for a spinning BH $\alpha$
by hand to account for the gravitational redshifts.\footnote{In
  Boyer-Lindquist coordinates for a Kerr BH,
\begin{eqnarray}
\alpha &=& \frac{\rho}{\Sigma}\sqrt{\Delta} \\ \rho &=& \left (\rbh^2
+S^2\cos^2\theta \right )^{1/2} \nonumber \\ \Sigma &=& \left (\left
[\rbh^2+S^2\right ]-S^2\Delta \sin^2\theta \right )^2 \nonumber \quad
\quad .
\end{eqnarray}
for BH spin $S\le 1$. Here we use $\rbh$ for the distance from the BH
to be distinguished from the distance from the neutron star $r$.}
Given a dipole magnetic field, which drops off with distance from the
NS as $r^{-3}$, anchored on the NS with radius $\RNS$ (taken to be 10 km throughout) and surface
magnetic-field strength $B_{\rm NS}$,
\begin{equation}
B(r)=B_{\rm NS}\left (\frac{\RNS}{\rns}\right )^3,
\label{Eq:BNS}
\end{equation}
the voltage (\ref{Eq:Volts}) acquires a contribution only from the
integral along the horizon in the direction of the line connecting the
BH and NS, and so evaluates to
\begin{align}
\label{HVolt}
V_{\mathcal{H}} = 2 R_H \left[ \frac{\Rbin \left (\Omega_{\orb} - \Omega_{\rm
    NS} \right)}{c} + \frac{S}{4\sqrt{2}} \right] B_{\rm NS}
\left(\frac{\RNS}{\Rbin} \right)^3,
\end{align} 
where $R_{H}$ is the radius of the horizon and where we have included
a factor to account for the spin, $0 \leq S \leq 1$, of the BH
\cite{McL:2011}. Notice that in Eq. (\ref{Eq:BNS}), $B_{\rm{NS}}$
drops off with distance from the NS, so the voltage varies across the
horizon for small binary separations. In the limit in which we ignore
the finite size of the compact objects, we interpret $r$ as the binary
separation.

The total power that can be liberated by the battery is
\begin{align}
\label{CircuitPower}
\Lum(t) = \frac{ V^2_{\mathcal{H}}(t) }{( \Res_{\mathcal{H}} +
  \Res_{\rm NS} )^2 } \Res_{\rm NS} .
\end{align}
The resistance across the horizon of the BH is
$\Res_{\mathcal{H}}=4\pi/c$ cm$^{-1}$s.  Since the effective
resistance of the NS and its magnetosphere ($\Res_{\rm NS}$) is
unknown, we choose $\Res_{\rm NS} = \Res_{\mathcal{H}}$ to give the
largest possible luminosities. This impedance matching condition is
the same as that imposed to derive the Blandford-Znajek power
\citep{BZ:1977}, in which case the angular velocity of magnetic-field
lines at infinity are set to one half of the BH horizon angular
velocity \citep{MPBook, Penna:ImpMatch2015}.

The power scales roughly as
\begin{equation}
\Lum \sim M^2 B_{\rm NS}^2 \Rbin^{-6} v^{2} \ .
\end{equation}
At large separations $v^2 \sim M/\Rbin$ is small, climbing to near the
speed of light at merger.  Measuring length in units of $M$, the power
scales as
\begin{equation}
\Lum \sim B_{\rm NS}^2 M^{-4} v^{2} \ .
\end{equation}
For a fixed number of gravitational radii between the NS surface and
the BH horizon, a larger BH boosts the power as $M^2$, but the larger
implied distance between the two decreases the magnetic-field strength
at the horizon by $M^{-6}$.

We discuss briefly when these scalings break down. In the limit that
the NS and BH are close, and their finite sizes are important, the NS
surface can come arbitrarily close to the BH horizon in which case
$B_{\rm NS}^2 \Rbin^{-6} \rightarrow B_{\rm NS}^2$. Placing the NS
surface at the horizon and spinning it with velocity $v$ would
generate power which increases with BH mass as $\Lum \sim M^2 B_{\rm
  NS}^2 v^{2}$. If however, the BH mass was very large, the variation
of the magnetic field across the BH horizon would become
important. For very large BHs, the NS light cylinder will not span the horizon.\footnote{When the BH event horizon is larger than the size of the NS light cylinder, $M \gtrsim c^3 G^{-1} \Omega^{-1}_{\rm
  NS} \sim 10^4 M_{\odot} 2 \pi/\Omega_{\rm NS}$, the full voltage drop of Eq. (\ref{HVolt}) cannot be realized.}
In these cases, our assumption that the voltage drop is across the entire horizon breaks down and the
power will scale more weakly than $M^2$. In the present work, we
ignore finite-size effects and take
Eqs. (\ref{Eq:BNS})-(\ref{CircuitPower}) to be a good estimate of the
average power available via the BH battery.

Here and throughout the rest of the paper we treat the NS surface magnetic-field strength as an unknown parameter. Because there are no observations of BHNS binaries, and hence no measurements of NS field strengths near merger with a BH, we have chosen a range in accordance with the observed NS fields \citep[see \textit{e.g.}][]{Kaspi:2016}. We consider fields ranging from those of the radio pulsar population $10^{12}$ G up to the observed magnetar field strengths of a few times $10^{15}$G \citep{Magnetars:2014} and beyond to larger, but not impossible field strengths of $10^{16}$G,\footnote{NS field strengths as high as $\sim10^{18}$G are theoretically possible but would generate EM power that would rival the emission due to gravitational radiation and hence require numerical analysis.} in order to probe the full range of energies available to the BHNS system. 
Conversely and as we discuss in \S \ref{Observability}, our models can constrain the NS field strength at merger.

\begin{figure}
\begin{center}
\includegraphics[scale=0.36]{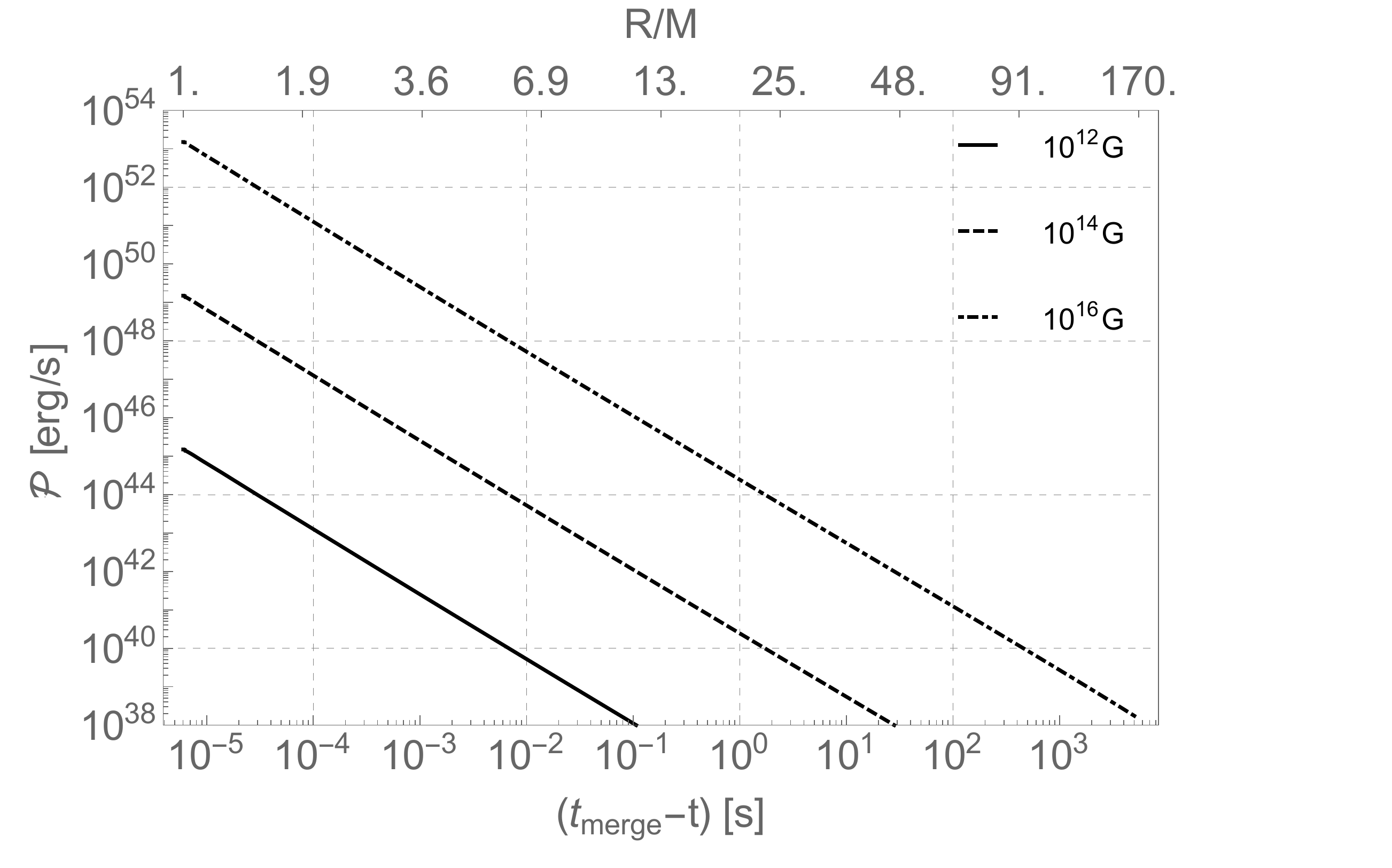}
\end{center}
\caption{Total possible power supplied by the BH battery via
  Eq.\ (\ref{CircuitPower}) as a function of time until merger for two
  point masses undergoing orbital decay via gravitational radiation
  reaction (Eq. \ref{PetersSep}). The solid, dashed, and dot-dashed
  lines indicate NS surface magnetic-field strengths of $10^{12}$,
  $10^{14}$, and $10^{16}$G respectively, for a BH mass of $10
  \Msun$. The plot extends to a binary separation of $G\MBH/c^2$, the
  size scale of the event horizon for the maximally spinning BH we
  consider. We hove dropped factors of $G$ and $c$ in the axis
  labels.}
\label{Fig:Lum_B_M}
\end{figure}

In Figure \ref{Fig:Lum_B_M}, we plot the total power available for
liberation by the binary as a function of time for varying NS magnetic
field strengths and a maximally spinning BH of mass $10M_\odot$.\footnote{Depending on the NS equation of state, the choice of a maximally spinning BH could cause the NS to be partially disrupted \citep[\textit{e.g.}][]{Foucart:2012}. In the same study, a BH spin $S\lesssim0.95$ does not disrupt, and changing the spin by such a small amount has no notable impact on our results.}
Importantly, over the range of possible magnetic-field strengths, 
the energy liberated through the BH-battery mechanism is
many orders of magnitude lower than that liberated by gravitational
radiation \citep{McL:2011}, hence the orbital inspiral time scales are
set by gravitational radiation loss and are robust despite different
possible channels for the electromagnetic power.  The time-dependent
separation $\Rbin(t)$ decays due to gravitational radiation losses
\cite{Peters64},
\begin{align}
\label{PetersSep}
\Rbin(t) = \left( \Rbin^4(0) - 4 \ \frac{64}{5} \frac{G^3}{c^5} M_{\rm NS} M \left(M + M_{\rm NS}\right) t \right)^{1/4},
\end{align}
where $M_{\rm NS}$ is the NS mass taken to be $1.4 \Msun$ throughout.
Over the final second, the power available climbs by $\sim8$ orders of
magnitude. For a $10^{12}G$ dipole field, the power rises from pulsar
scales $\sim 10^{36} {\rm erg \ s}^{-1}$ in that second, to
$\sim10^{44} {\rm erg \ s}^{-1}$ in the final millisecond (at $\Rbin=2
G \MBH/c^2$).  The power scales as $B^2$ reaching $10^{52} {\rm erg
  \ s}^{-1}$ for a magnetar with $B\sim10^{16}G$. For a maximally
spinning BH, the horizon is at $\Rbin=G \MBH/c^2$, so we extend the
luminosity scaling in Figure \ref{Fig:Lum_B_M} down to this separation
(noting that we still have $G\MBH/c^2 > R_{\rm{NS}}$ for $\MBH \geq 7
\Msun$) where the luminosity peaks at $\sim 10^{45} {\rm erg \ s}^{-1}
(B/10^{12} \rm{G})^2$.

%%%%%%%%%%%%%%%%%%%%%%%%%%%%%%%%%%
%%%FIGURE Lum vs t
%%%%%%%%%%%%%%%%%%%%%%%%%%%%%%%%%%
%%%%%%%%%%%%%%%%%%%%%%%%%%%%%%%%%%%%%%%%%%%%%%%%

Equation (\ref{CircuitPower}) gives an estimate of the power the battery could
generate. Whether or not this power is available to light up the pair
is the question at hand. We describe the most straightforward vehicles
to convert the power into luminosity in the following sections.

\section{Curvature radiation}
The voltage drop will accelerate charges across magnetic-field lines
connecting the NS to the BH. Basic physics suggests that these
accelerated charges will provide a sensible channel for luminosity.
The charges spiral around and are pushed along the magnetic fields
when there is a parallel component of electric field, $\Evec \cdot
\Bvec \ne 0$.  The result is a primary spectrum of curvature
radiation.\footnote{When the energy of curvature photons is great
  enough, they will interact with the magnetosphere magnetic and
  electric fields and produce electron-positron 
  pairs. As the curvature photons are not locked to move along
  magnetic-field lines, the secondary pairs can have a non-negligible
  component of motion transverse to the magnetic field, resulting in a
  secondary synchrotron spectrum.}

The extent to which the BH battery can act as a particle accelerator
is mitigated by the conducting properties of the surrounding
magnetosphere. The NS sustains a magnetosphere by pulling charges from
the NS and through various pair production channels in the
magnetosphere \citep{GJ:1969, RudSuth:1975}.  The plasma acts as a
conductor and will screen the NS's electric fields until force-free
conditions are established, that is, until $\Evec \cdot \Bvec =0$.

Once the BH enters the light cylinder of the NS and the battery is
established, the electric field configuration changes and the
magnetosphere adjusts with those changes. At the large separations of
the light cylinder, the plasma is tenuous but in the final stages when
the voltage is most powerful, both compact objects should be submerged
in the conducting plasma. Consequently, we anticipate that some of the
emf generated by the orbital motion is screened and forces are
muted. However, as with the pulsar, there must be gaps in which
screening is inefficient and across which particles must be
accelerated. Additionally, current sheets could act to dissipate the
BH-battery power.

We currently do not know the degree to which the voltage is reduced by
screening. In the future, global particle-in-cell codes could asses
the gap structure in a BHNS magnetosphere. To make simple estimates,
we continue to use the full power of the battery in the calculation of
the curvature radiation, aware that screening could significantly
reduce the estimates.

To obtain the primary curvature radiation spectrum, we assume a
distribution in energy of the magnetosphere electrons and
positrons. The spectrum of curvature radiation is given by integrating
the one-electron spectrum multiplied by the number distribution of
charged particles.
\begin{equation}
\label{PofOmega}
P_C(\nu,t) = \int^{\gamma_{\rm max}}_{\gamma_{\rm min}}{ N(\gamma)
  \frac{dP_{C}}{ d \nu} d\gamma}
\end{equation}
where $dP_C/d\nu$ represents the curvature radiation power per unit
frequency \citep[\textit{e.g},][]{CZ_SC:1996}.  We model the
population as a power law in the relativistic Lorentz factor $\gamma$,
          \begin{equation}
           \label{New:PowerLawElecs}
          N(\gamma) d\gamma = N_0 \gamma^{-p} d\gamma.
          \end{equation} 
The normalization constant $N_0$ is chosen so that the total
bolometric luminosity matches Eq. (\ref{CircuitPower})
\begin{equation}
N_0 = \frac{ \Lum }{ \int{ \int{ \gamma^{-p} \frac{dP_{C}}{ d \nu} d\gamma} \ d\nu} },
\end{equation}
so that the magnetosphere number density ($\sim N_0/\Rbin^3$) is set
by the physics of curvature radiation and the requirement that the
magnetosphere maximally radiates the BH-battery power.

The spectrum then depends on the energy distribution of electrons and
positrons through the exponent $p$, and the time-dependent minimum and
maximum Lorentz factors of particles in the magnetosphere $\gamma_{\rm
  max}(t)$ and $\gamma_{\rm min}(t)$ that we must input from the
physical model of the BHNS battery. As the spectrum is not greatly
dependent on the minimum $\gamma$ or the power law index $p$ (see the
Appendix), we leave these as free parameters. The shape of the
spectrum will depend on the choice of $N(\gamma)$, but, for what
follows, the most important consideration will be where the high
energy end of the spectrum is cut off. This is set by the maximum
electron Lorentz factor in the magnetosphere.

We approximate the maximum $\gamma$ as the largest radiation-reaction
limited Lorentz factor in the magnetosphere. Electrons and positrons
are accelerated along magnetic-field lines to radiation-reaction
limited velocities given by solving,
\begin{align}
 e c |\Evec_{||}| \left( 1 - \gamma^{-2}\right)^{1/2}_{\rm max}=
 \frac{2}{3} \frac{c e^2 \gamma^4_{\rm max}}{ \rho_c^2}
 \label{Eq:GamMax}
\end{align}
for the Lorentz factor $\gamma_{\rm max}$. Here $\rho_c$ is the radius
of curvature of magnetic-field lines. We evaluate $\rho_c$ for a
dipole magnetic field in the binary equatorial plane, $\rho_c = \RNS/3
\sqrt{r/\RNS}$.  We use the horizon electric field sourced by the
potential drop Eq. (\ref{HVolt}) to estimate a maximum value of the
accelerating electric fields, $ |\Evec_{||}| \approx |\Evec| \sim
\frac{V_{\mathcal{H}}}{ R_{H}} $ where $R_H$ is
the radius of the BH horizon.

Then the radiation-reaction limited Lorentz factor of
electrons/positrons, at the BH horizon is
\begin{align}
\gamma_{\rm max} &\approx 4.2 \times 10^7 \left(\frac{\Rbin}{6GM/c^2}
\right)^{-5/8} \left(\frac{B_{\rm NS}}{10^{12} G} \right)^{1/4},
\label{Eq:GamMax}
\end{align}
choosing fiducial parameters $\RNS = 10^6$ cm and $M_{\rm BH}= 10
{\Msun}$.  Electrons and positrons will emit curvature radiation with
characteristic energy
\begin{align}
\epsilon_{\gamma} = \frac{3hc}{4 \pi \rho_c} \gamma^3 \approx 1.8
\ \mbox{TeV} \left(\frac{\gamma}{4.2 \times 10^7}\right)^3.
\end{align}
We plot a representative curvature radiation spectrum for a fiducial
$10 \Msun$ BH with maximal spin. The dependence of the curvature
spectrum on $\gamma_{\rm min}$ and $p$ is explored in the Appendix.

\begin{figure}
\begin{center}$
\begin{array}{c c}
\includegraphics[scale=0.32]{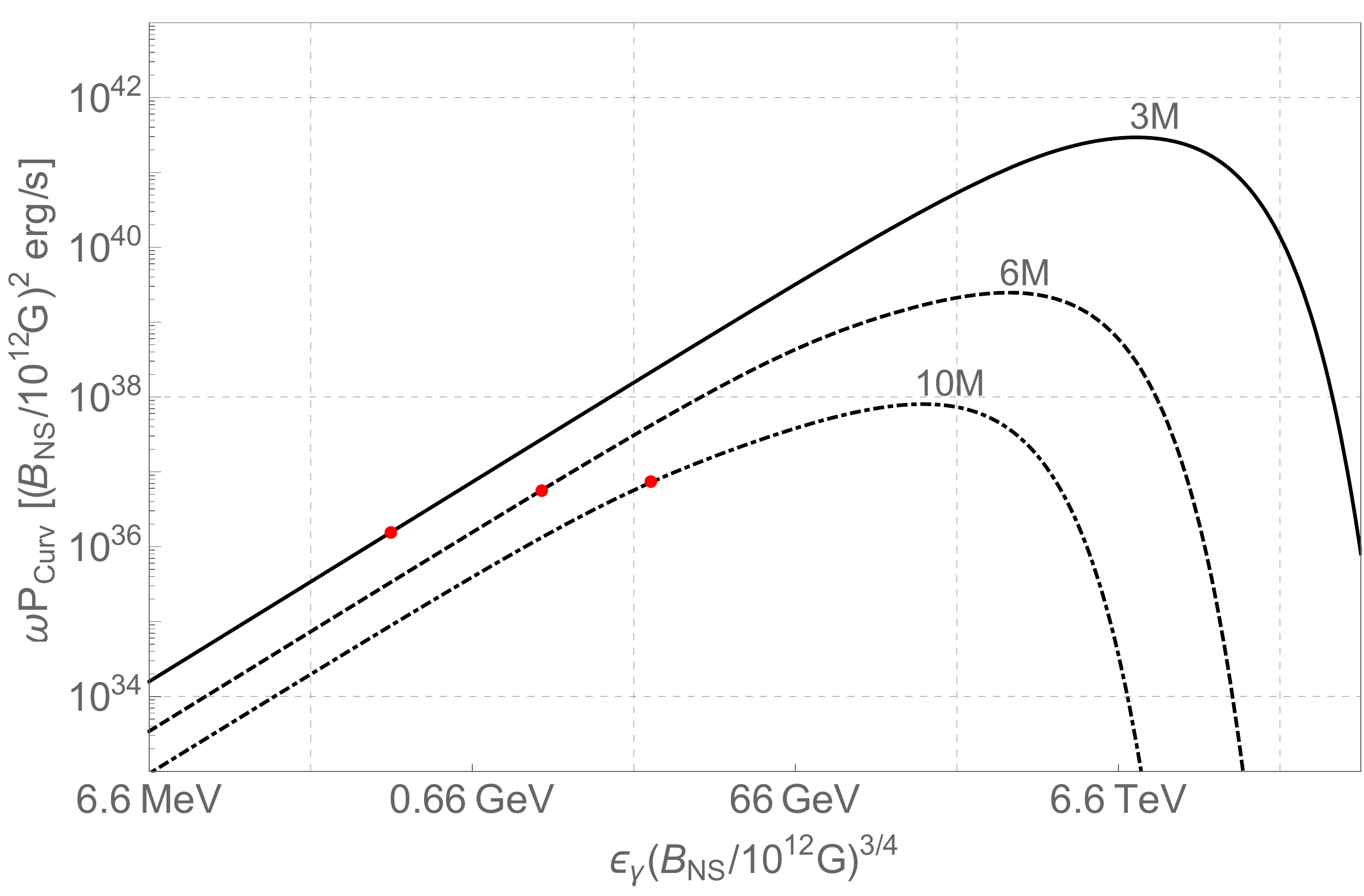}
\end{array}$
\end{center}
\caption{The spectra of primary curvature radiation at times
  corresponding to binary separations $10GM/c^2$, $6GM/c^2$, and
  $3GM/c^2$ (dot-dashed, dashed, solid) scaled to $B_{\rm
    NS}=10^{12}G$ (factors of $G$ and $c$ are omitted in the
  labels). We use an electron-energy power law index of $p=2.0$ and a
  minimum Lorentz factor set by radiation reaction in the outer
  magnetosphere. Dependence on both parameters is minimal (see the
  Appendix). The red dots indicate photon energies above which the
  magnetosphere is opaque to pair production via $\gamma + B$
  interactions.}
\label{Fig:CurvSpectra_p2}
\end{figure}

In agreement with previous works \citep{McL:2011, DorazioLevin:2013},
Figure \ref{Fig:CurvSpectra_p2} shows that the BHNS curvature
radiation can be very high energy, $>$TeV, near merger.  In the
following section, we point out that this curvature radiation will be
prone to copious pair production through interaction with the strong
electromagnetic fields of the magnetosphere as well as photon-photon
collisions. The pair production will further populate the
electron-positron plasma surrounding the binary. Depending on the
efficiency at which pairs are produced from the available energy of
the BH battery, the magnetosphere will become optically thick to
curvature photons. This trapped radiation can power a fireball, which
we now characterize.

%%%%%%%%%%%%%%%%%%%%%%%%%%%%%%%%%%
%%%FIGURE Curv and Synch Spectra 
%%%%%%%%%%%%%%%%%%%%%%%%%%%%%%%%%%
%%%%%%%%%%%%%%%%%%%%%%%%%%%%%%%%%%%%%%%%%%%%%%%%

\section{Fireball}
\label{Fireball}
As the BH and NS draw closer, the energy available to accelerate
particles increases as $\Rbin^{-3}v$, resulting in a higher density of
higher energy curvature photons. A consequence is pair production
through the interaction of the magnetic field and high-energy photons
($\gamma + B \rightarrow e^{+} + e^{-}$) and through photon collisions
($\gamma + \gamma \rightarrow e^{+} + e^{-}$), preventing the highest
energy curvature photons from escaping the magnetosphere.  The result
is an optically thick pair+radiation fluid, which will expand outwards
under its own pressure until pair production becomes disfavored and
radiation can escape; the result is a fireball.

\subsection{Pair production}
The optical depth to $\gamma + B \rightarrow e^{+} + e^{-}$, at binary
separation $\Rbin$ is
\begin{align}
\label{GBoptd}
\tau_{\gamma B} &= \Rbin \left[\frac{4.4}{e^2 / (\hbar c)}
  \frac{\hbar}{m_e c} \frac{B_q}{B_{\perp}} \exp{\left(\frac{4}{3 \xi}\right)} \right]^{-1}  \\ 
\xi &\equiv \frac{\hbar \omega}{2 m_e c^2} \frac{B_\perp}{B_q} \nonumber \\ 
B_q & \equiv \frac{m^2_e c^3}{e \hbar} \approx 4.4 \times 10^{13} \rm{G} \nonumber \\ \nonumber
   B_{\perp} &\equiv \rm{Min}\left\{x/(\RNS/ 3\sqrt{r/\RNS}),  1 \right\} B(r)
\end{align}
for photons with $\hbar \omega \gtrsim 2 m_e c^2$. The quantity in
brackets is the mean free path for pair production given by
Refs. \cite{Erber:1966, RudSuth:1975}, $B_q$ is a natural quantum
mechanical measure of magnetic-field strength, and $B_{\perp}$ is the
component of magnetic field perpendicular to the photon trajectory. 
The quantity in curly brackets in the last line of Eq. (\ref{GBoptd}) is the 
sine of the angle between a photon trajectory and the magnetic-field 
direction, which is simply the distance $x$ a photon has traveled in direction
initially tangent to a field line, divided by the radius 
of curvature of field lines. As a characteristic value, we take the radius 
of curvature to be that of a dipole field line which goes through the center 
of the BH at binary separation $r$.
This approximation assumes that $\xi \ll 1$, which is always true initially when $x=0$
and $B_{\perp} = 0$. In practice we cap $\xi\leq 1$ because we are
only interested in when $\tau_{\gamma B} \rightarrow 1$.  After this
point the $\gamma + \gamma \rightarrow e^{+} + e^{-}$ process will
also become important, so we need not rely solely on the above
calculation (see below).

For very high-energy photons, the optical depth limits to very large
values but drops exponentially for lower energy photons, generated
earlier in the binary inspiral. To capture the steep dependence of the
$\gamma + B \rightarrow e^{+} + e^{-}$ optical depth on photon
frequency, we evaluate $\tau_{\gamma B}$ at a frequency near the peak
of the time-dependent curvature radiation spectrum (see Figure
\ref{Fig:CurvSpectra_p2}).

\begin{figure}
\begin{center} \vspace{-20pt}
\includegraphics[scale=0.38]{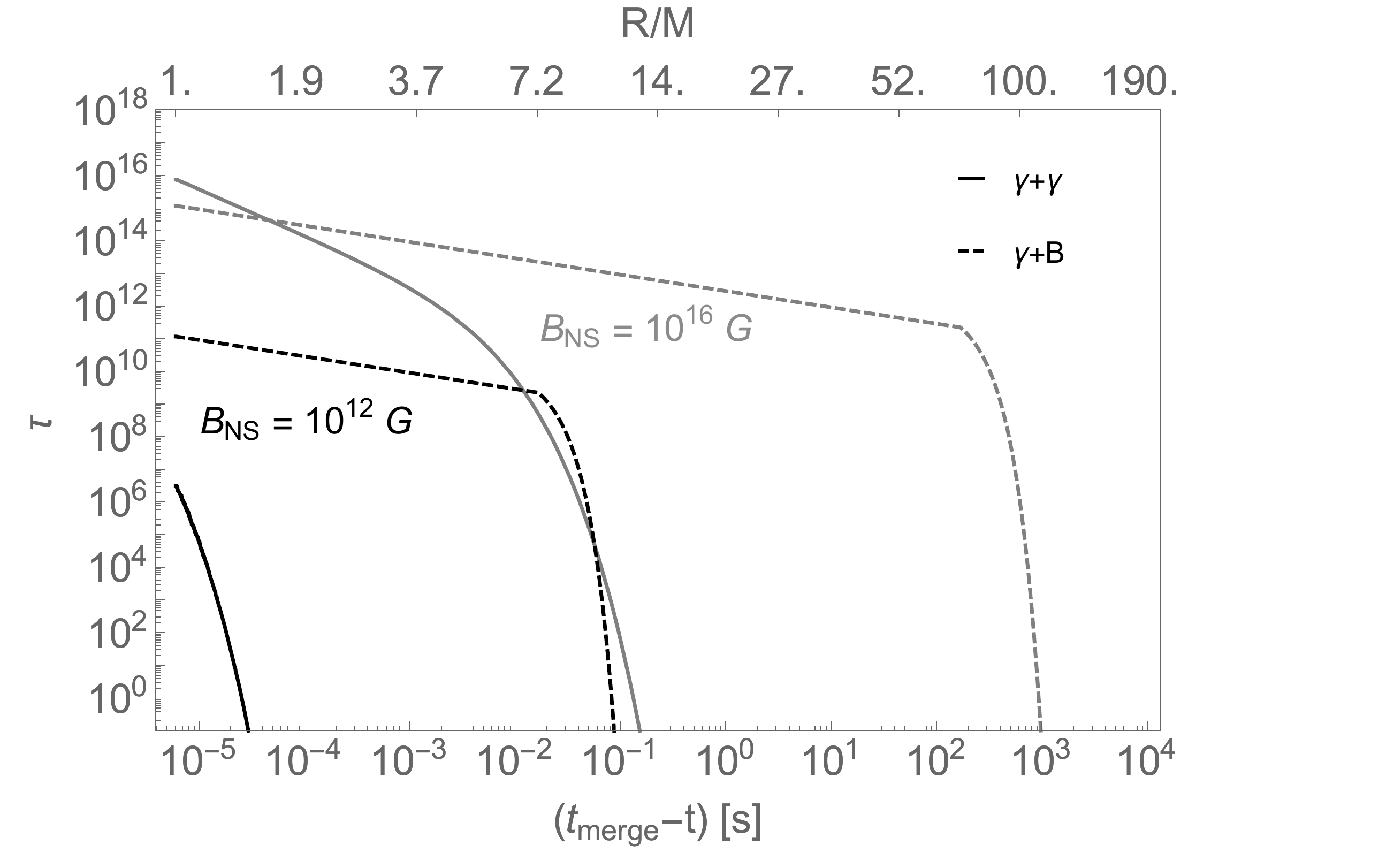}
\end{center}
\caption{The optical depth to the different pair producing
  processes. The magnetosphere curvature photons are trapped by
  $\gamma + B$ early on, $\gamma + \gamma$ also becomes relevant for
  magnetosphere photons just before merger. The $\gamma+B$ optical
  depth is computed at a time-dependent frequency near the peak of the
  primary curvature spectrum. Factors of $G$ and $c$ are omitted in
  the upper x-axis label.}
\label{Fig:tauComp}
\end{figure}

The red dots plotted on top of the spectra of Figure
\ref{Fig:CurvSpectra_p2} show the frequency at which the $\gamma + B
\rightarrow e^{+} + e^{-}$ optical depth (Figure \ref{Fig:tauComp})
becomes unity for three different snapshots during the inspiral. Above
the frequency indicated by the red dots in Figure
\ref{Fig:CurvSpectra_p2}, photons pair produce with the magnetic field
before escaping the magnetosphere.

The optical depth for $\gamma + \gamma \rightarrow e^{+} + e^{-}$ at
binary separation $\Rbin$ is
\begin{equation}
\label{GGoptd}
\tau_{\gamma \gamma} \approx \Rbin n_{\gamma*} \sigma_{\gamma \gamma}
\end{equation}
where we use a collision cross section $\sigma_{\gamma \gamma} =
11/180 \sigma_T$ \citep{LithSari:2001, Sven:1987} averaged over photon
energy and written in terms of the Thomson scattering cross section
$\sigma_T$.

Once the magnetosphere becomes optically thick to $\gamma + B$ pair
production, we assume that the radiation plus pair plasma thermalizes.
Then we may approximate $n_{\gamma*}$ as the portion of the Planck
spectrum with sufficient energy to produce pairs
\begin{equation}
\label{nstar}
n_{\gamma*} = \frac{8 \pi }{c^3} \int^{\infty}_{2 m_e c^2/h}{\frac{
    \nu^2 \ d\nu}{ \mbox{e}^{h \nu /kT} -1 } } .
\end{equation}
which is an underestimate as any two photons with energies
$\sqrt{\epsilon_1 \epsilon_2} \geq 2 m_e c^2$ are favored to create
pairs upon collision, not just those above $2m_e c^2$.

%%%%%%%%%%%%%%%%%%%%%%%%%%%%%%%%%%
%%%FIGURE tau gamgam  and tau gamB vs t
%%%%%%%%%%%%%%%%%%%%%%%%%%%%%%%%%%
%%%%%%%%%%%%%%%%%%%%%%%%%%%%%%%%%%%%%%%%%%%%%%%%

%%%%%%%%%%%%%%%%%%%%%%%%%%%%%%%%%%
%%%FIGURE R_photosphere
%%%%%%%%%%%%%%%%%%%%%%%%%%%%%%%%%%
\begin{figure}
\begin{center}
\includegraphics[scale=0.32]{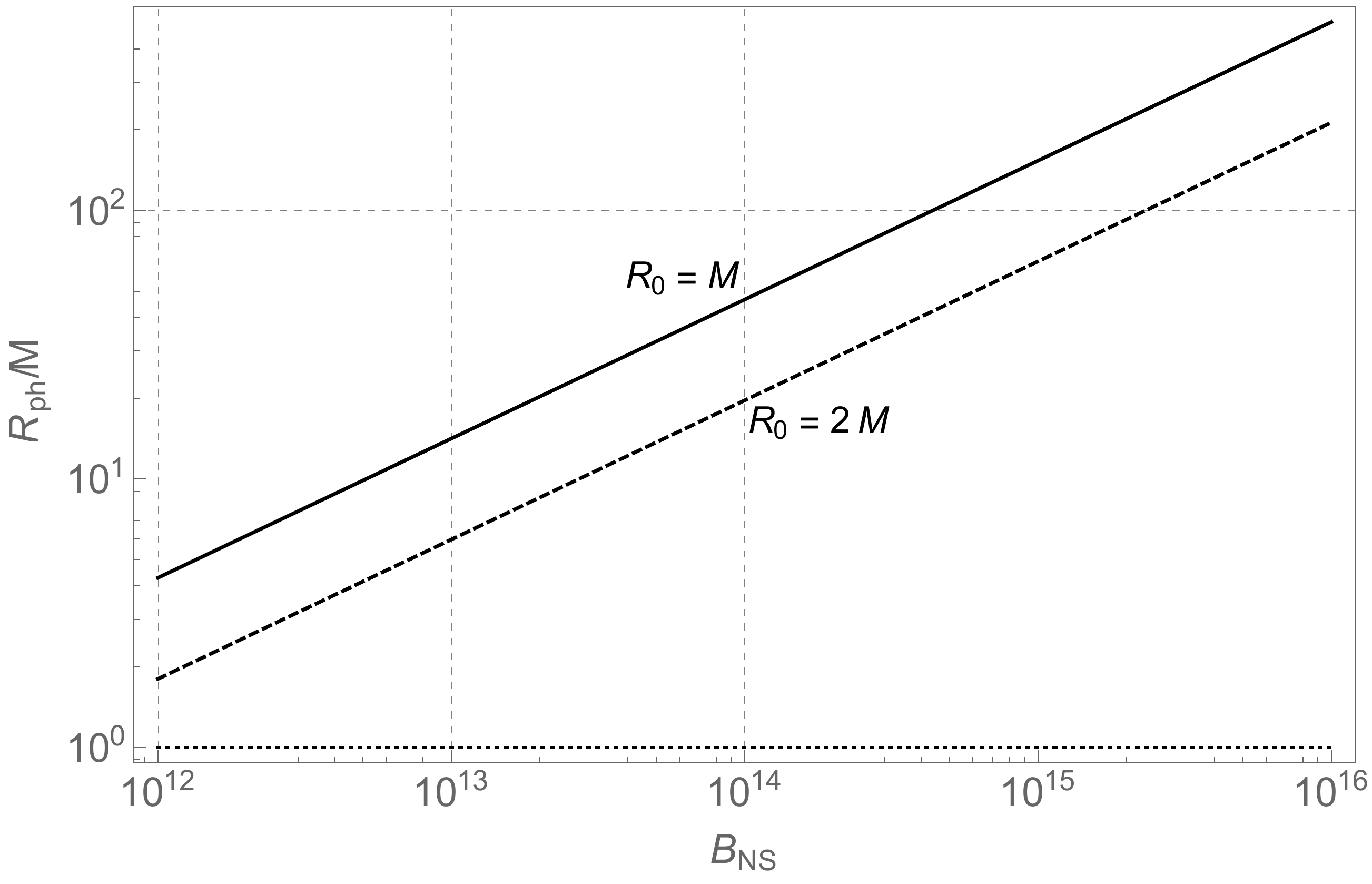}
\end{center}
\caption{The radius of the photosphere as a function of NS
  magnetic-field strength, for different assumed radii of energy
  injection $R_0=GM/c^2, 2GM/c^2$. Factors of $G$ and $c$ are omitted
  in the figure labels.}
\label{Fig:Rphot}
\end{figure}
%%%%%%%%%%%%%%%%%%%%%%%%%%%%%%%%%%%%%%%%%%%%%%%%

Figure \ref{Fig:tauComp} shows the optical depth of the magnetosphere
to both $\gamma + B$ and $\gamma + \gamma$ pair production as a
function of time during inspiral for NS magnetic-field strengths which
bracket the expected range. The $\gamma + B$ process becomes important
first, when curvature-photon energies surpass a critical value (see
the red dots plotted on the spectra of Figure
\ref{Fig:CurvSpectra_p2}). Much closer to merger, $\gamma + \gamma
\rightarrow e^- + e^+$ also becomes an important source of pair
production and hence photon opacity.

The high optical depths in Figure \ref{Fig:tauComp} suggest copious
pair production due to $\gamma + B$ earlier in the inspiral.  If this
process thermalizes the radiation and pairs, then our assumption of a
Planck gas in the computation of the subsequent $\gamma + \gamma$
optical depth is warranted. The important point is that, with the
large magnetic-field strengths and energy densities present in the
BHNS magnetosphere near merger, both pair production processes will be
favored. Hence we reason that pair production traps and thermalizes
the power generated by the BH battery.

We can conclude from this section that the era of curvature radiation
gives way to a hot fireball in the final moments before
merger. Curvature radiation becomes trapped when $\tau_{\gamma B}=1$
(Figure \ref{Fig:tauComp}), from which we find that high-energy
curvature radiation will no longer escape for the final $0.1$s
$(B/10^{12}\rm{G})$ of inspiral. Figure \ref{Fig:Lum_B_M} shows that at 
$0.1$s $(B/10^{12}\rm{G})$ before merger the BH-battery luminosity, 
and thus the maximum power in curvature radiation, is $\sim 10^{38} \rm{erg} \ \rm{s}^{-1}
\left(B/10^{12}\rm{G}\right)^{1/2}$, a factor of $\sim 10^7
\left(B/10^{12}\rm{G}\right)^{3/2}$ lower than the BH-battery peak
power at merger. Consequently, at $P_{C} \lesssim 10^{38} \rm{erg} \ \rm{s}^{-1}
\left(B/10^{12}\rm{G}\right)^{1/2}$, the ramp up in high-energy curvature radiation
will likely only be observable within the galaxy. 

The subsequent
fireball however, could be observable at cosmological distances. We
characterize the emission from the fireball in the following section.

\subsection{Expansion and emission}
The optically thick pair plus radiation fluid -- the fireball -- will
expand under it's own pressure. The alternative is that the fireball
falls right down into the BH, although we argue this will not happen.  To
determine if the fireball will expand, we consider the imbalance of
gravity and the mechanical pressure $P$ of the fluid. The condition
for expansion is
\begin{equation}
\frac{dP}{dr}\sim \frac{P}{R_0} > \rho \frac{GM}{R_0^2},
\label{Eq:PGbal}
\end{equation}
where $R_0$ is the initial scale over which energy is injected by the
battery. For a radiation dominated fluid $P=\rho c^2/3$ and then
\begin{equation}
R_0 \gtrsim  \frac{GM}{c^2},
\end{equation}
dropping all numerical factors. Radiation pressure alone can cause 
the fireball to expand. We note that the force balance is marginal at 
small size scales and will depend on the density distribution in addition 
to magnetic pressure, both of which will likely increase the outward 
pressure of the fireball and should be treated in a more detailed 
calculation. Considering the high temperature at merger, the pressure 
may be dominated by pairs, not radiation. In this limit, $kT > m_e c^2$, 
the total pair pressure is $7/4$ the radiation pressure and the fireball will still expand.

After merger, the magnetic fields responsible for $\gamma + B$ pair
production will decay without the NS to anchor them (see however \S
\ref{Post-Merger}). This means that, after merger, only $\gamma +
\gamma$ pair production and electron scattering will trap photons in
the expanding fireball. To track the expansion of the fluid from this
point, we estimate its properties during and after merger.

Because the optically thick, pair plus radiation fluid is assumed to
be in thermal equilibrium, we can estimate the temperature of the
fluid as
\begin{align}
T(\Rbin) &=  \left( \frac{ \Lum(\Rbin) }{ 4 \pi \Rbin^2 \sigma } \right)^{1/4}, \nonumber \\
\end{align}
as a function of the binary separation throughout inspiral, where
$\Lum(\Rbin)$ is the power emitted by the BH battery at separation
$\Rbin$, and $\sigma$ is the Stefan-Boltzmann constant. Then the
initial temperature of the fireball $T_0$ is the final temperature
before the magnetic fields are swallowed/dissipated and the pair plus
radiation fluid is released to expand. Evaluating this temperature at
a final binary separation of $R_0 \sim GM/c^2$ gives an initial
injection temperature of
\begin{align}
kT_0= 85 \ \mbox{keV} \left( \frac{B_{\rm NS}}{10^{12} G} \right)^{1/2} .
\label{Eq:KTeff}
\end{align}

We treat the fireball as an adiabatically expanding, relativistic
fluid. As the fluid expands to a radial size scale $R$, it cools as $T
= T_0 (R/R_0)^{-1}$. At a large enough $R$, $\gamma + \gamma$ pair
production and electron scattering will no longer trap photons, and
radiation escapes.

The $\gamma + \gamma$ optical depth is given by Eq. (\ref{GGoptd}) and
the optical depth to electron/positron scattering is,
\begin{equation}
\tau_{\rm{es}} \sim \Rbin n_{\pm} \sigma_{\rm{T}}, 
\end{equation}
where $\sigma_{\rm{T}}$ is the Thomson scattering cross section, and
$n_{\pm}$ is the rest-frame, pair number density in thermal
equilibrium. We estimate $n_{\pm}$ as the electron number density
\citep[{\em e.g.},][]{Pacz:1986GRB}, true for $kT \ll m_e c^2$, which
is always the case in the photosphere for $B_{\rm{NS}} \lesssim 10^{16}$
G. Then,
\begin{align}
\label{Eq:npm}
n_{\pm} &\approx \frac{4 \pi^{3/2}}{h^3} (2 m_e k T)^{3/2} \rm{exp}\left(- \frac{m_e c^2}{k T}\right) 
\end{align}
Eventually the fireball expands until the temperature has dropped
sufficiently for both $\tau_{\gamma \gamma} \leq 1$ and $\tau_{\rm es}
\leq1$. We call this radius the photosphere radius $R_{\rm{ph}}$. We
find that the fireball first becomes transparent to $\gamma + \gamma$
pair production and then to electron scattering at a larger, but
similar radius (within a factor of a few). Hence the photosphere is
defined where $\tau_{\rm es}(R_{\rm{ph}}) \equiv 1$. The photosphere
radius as a function of NS magnetic-field strength is plotted in
Figure \ref{Fig:Rphot} for two choices of the initial size of the
fireball, $GM/c^2$ and $2GM/c^2$ (we assume a fiducial $R_0=GM/c^2$
throughout).

We estimate the Lorentz factor of the adiabatically expanding fluid as
$\gamma = R/R_0$ \citep{Pacz:1986GRB} for $R\gg R_0$.  Then emission
from the photosphere will be that of a blackbody boosted at Lorentz
factor $\gamma_{\rm{ph}} = R_{\rm{ph}}/R_0$. Such a boosted blackbody
looks like the rest-frame blackbody but with an effective temperature
\be 
T_{\rm eff} = \frac{T_{\rm{ph}}}{\gamma_{\rm{ph}} (1-v_{||}/c) } \equiv D T_{\rm{ph}}
\label{Eq:Teff}
\ee where $D$ is the doppler factor, $T_{\rm{ph}}$ is the temperature
in the rest frame of the photosphere, and $v_{||} = v\cos{\theta}$ is
the line-of-sight velocity, where $\theta$ is the angle from observer
line of sight. Because the shell is expanding spherically, each patch
of the expanding photosphere will have a different effective
temperature and the observed, time-dependent spectrum will be a sum of
the spectra of all patches on equivalent light travel time surfaces
\citep[{\em e.g.},][]{Pe'er:2011}. We do not include such details
here; in \S \ref{Observability} we integrate the line-of-sight
dependent blackbody spectra over the photosphere to find a composite
spectrum, but for now we make a simple estimate for the peak energy of
blackbody emission.

The total photospheric emission will not deviate greatly from
blackbody, and the majority of emission will come from the portion of
the expanding sphere for which the Doppler factor is positive, where
the angle to the line of sight is less than $1/\gamma$. For highly
relativistic expansion, the blue-shifted temperature
Eq. (\ref{Eq:Teff}) becomes $T = \gamma T_{\rm{ph}}$ at
$\theta=1/\gamma$ and $T = 2 \gamma T_{\rm{ph}}$ at $\theta=0$. For
simplicity we use that the photosphere emission is a blackbody with
temperature $T \sim \gamma T_{\rm{ph}}$. Then because the photosphere
temperature is related to the initial temperature as $T_{\rm ph} =
T_0(R_0/R) = T_0/\gamma_{\rm{ph}}$, the observed blackbody temperature
is simply $T=T_0$ \citep[{\em see also}][]{Pacz:1986GRB}; the observed
temperature is the same as the initial injection temperature of
Eq. (\ref{Eq:KTeff}) (the effects of gravitational redshift are
negligible for $R_{\rm ph}\gg R_0$). For a fiducial energy-injection
size scale of $R_0 = GM/c^2$, the {\em observed} photosphere emission
will peak at
\begin{align}
h \nu_{\rm{peak}} = 0.24 \ \mbox{MeV} \left( \frac{B_{\rm NS}}{10^{12} G} \right)^{1/2} ,
\label{Eq:KTWein}
\end{align}
ranging from hard x rays to $\gamma$ rays.

From the pair density at the photosphere we estimate the plasma
frequency to be,
\begin{align}
\nu_{\rm pl} = \sqrt{\frac{ n_{\pm} e^2}{\pi m_e}} \lesssim 4.4 \times 10^{12} \rm{Hz}   \left( \frac{B_{\rm NS}}{10^{12} \rm{G}}\right)^{-0.26} .
\end{align}
The blackbody emission is not shorted out by the pair plasma, however,
emission in the far-infrared and at longer wavelengths does not escape
the photosphere.

Because the photosphere is generated due to a decrease in pair
density, there will be no detectable signal from blue-shifted pair
annihilation \citep[{\em see also}][]{Pacz:1986GRB,
  GoodmanGRB:1986}. The ratio of energy in pairs to that in radiation
at the photosphere is small,
\begin{align}
\frac{E_{\pm}}{E_{\gamma}} \simeq \frac{m_e c^2 n_{\pm} c}{\sigma T_{\rm{ph}}^4} < 10^{-8} \ .
\end{align}

Finally we note that, because the fireball must expand out to its
photosphere size before it can radiate, the EM transient predicted
here will occur at least $R_{\rm{ph}}/c \sim 0.2 \ \rm{msec}
\sqrt{B/10^{12}\rm{G}}$ after the initial energy injection. If energy
injection is associated with merger, then this EM signature will occur
shortly after peak gravitational-wave emission.  Hence gravitational 
waves from the inspiral stage, which will trigger a LIGO detection, 
will also warn of this EM counterpart.

To summarize, we predict that, as the binary nears the final few
$G\MBH/c^2$ in binary separation, high-energy curvature radiation will
produce pairs by interacting with other photons and also the magnetic
field. The BHNS magnetosphere becomes optically thick to pair
production, trapping the energy injected by the BH battery. This
energy injection causes the optically thick pair plus radiation fluid
to expand outwards until the temperature drops below that which favors
a high pair density. At this point pair production and electron
scattering no longer contain the photons and they escape. For initial
NS field strengths of $10^{12} \rightarrow 10^{16}$G, the observable
radiation is characterized as:
\begin{itemize}
\item Blackbody radiation with a peak photon energy $h \nu \sim$ 0.24
  MeV $\sqrt{B_{\rm NS}/10^{12}\rm{G}}$.
\item A bolometric luminosity of up to \\ $10^{45}$ erg s$^{-1}$
  $(B_{\rm NS}/10^{12}\rm{G})^2$.
\item Defining $\Delta t_{42}(B_{\rm NS})$ as the time before merger
  over which the BH is supplying power above $10^{42}$ erg s$^{-1}$,
  and associating this with the emission timescale, the the burst
  times to the closest order of magnitude are $\Delta t_{42}(10^{12}
  \rm{G}) \sim 10^{-3}$ s, $\Delta t_{42}(10^{14} \rm{G}) \sim 0.1$ s,
  $\Delta t_{42}(10^{16} \rm{G}) \sim 10$ s.
\end{itemize}
We next consider a post-merger signal and the observability of both
merger and post-merger events.

%%%%%%%%%%%%%%
%%Section: Post-Merger
%%%%%%%%%%%%%%
\section{Post Merger}
\label{Post-Merger}
When the BH swallows the NS, a magnetic flux is deposited onto the BH,
magnetizing the hole. The no-hair theorem suggests the BH, in vacuum,
must shed the absorbed $B$ field on order the BH light crossing time,
in very long-wavelength, $\sim R_{H}$, radiation \citep[{\em
    e.g.},][]{BaumShap:2003}. However, \cite{LyutikovMckinney:2011} have argued, in the context of NS collapse to a BH, that because the BH is immersed in magnetosphere plasma, the no-hair theorem is not applicable and the BH may retain a magnetic field anchored in a remnant magnetosphere for longer. The situation is similar to our case where the BH swallows the NS. In the limit of a nonresistive plasma, magnetic-field lines are frozen into the plasma of the magnetosphere. Because of the frozen-in condition, field lines which connect the NS surface to infinity before merger must also connect the BH horizon to infinity after merger, while closed field lines are swallowed along with the NS.  Hence a magnetic field is anchored onto the BH merger remnant. For a resistive plasma, the field will decay on the resistive timescale of the magnetosphere. As a consequence, the remnant BH could generate an electromagnetic signature through the BZ mechanism \citep{BZ:1977, LyutikovMckinney:2011}.

\begin{figure}
\begin{center}
\includegraphics[scale=0.33]{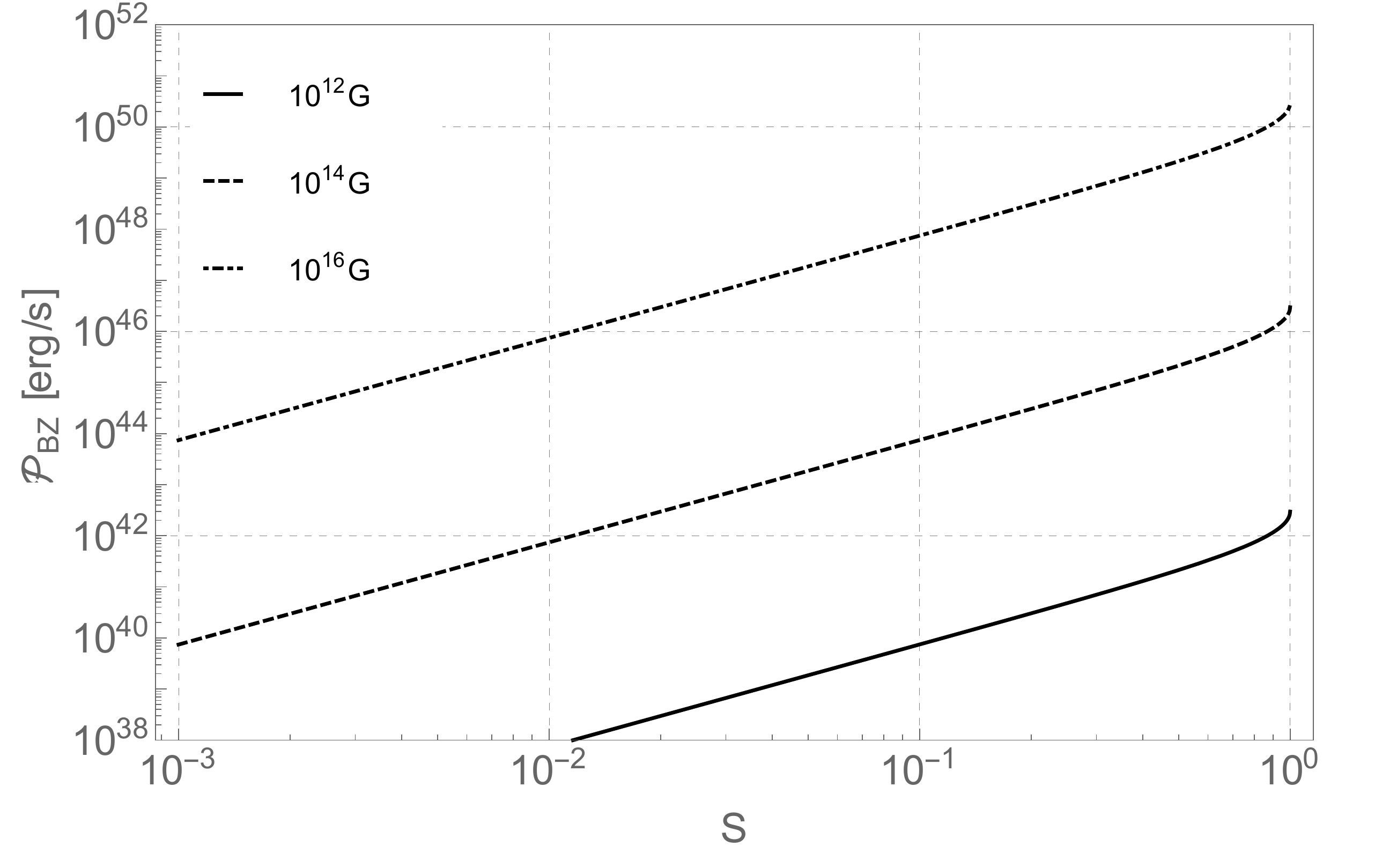} \vspace{-20 pt}
\end{center}
\caption{The power available to the post-merger, spinning BH remnant
  as a function of remnant spin and NS magnetic-field strength. This
  power is generated from the Blandford-Znajek process and the flux of
  open NS magnetic-field lines, Eq. (\ref{Eq:PBZ}). This maximal power
  will decay as the remnant magnetosphere decays on the resistive
  timescale.}
\label{Fig:PostMerger}
\end{figure}

The initial BZ power can be written in terms of the magnetic flux
deposited onto the BH horizon as
\begin{align}
  \label{Eq:PBZ}
&P_{\rm{BZ}} \sim \frac{\phi^2}{4 \pi c} \left(\frac{S
    c}{R_{H}(S)}\right)^2 \\ \nonumber 
&\sim 3 \times 10^{42}
  \rm{erg} \rm{s}^{-1} S^2 \left( \frac{B_{\rm{NS}} }{ 10^{12}
    G}\right)^2 \left( \frac{2 \pi / \Omega_{\rm{orb}} }{ 1 \rm{msec}}\right)^{-2}
  \left( \frac{R_H(S) }{ GM/c^2 }\right)^{-2} ,
\end{align}
where $S$ is the dimensionless BH spin related to the BH angular
momentum by $J = S GM^2/c$, $R_H(S)$ is the spin dependent horizon
radius, and $2 \pi / \Omega_{\rm{orb}}$ is the binary orbital period. In the second line we have approximated the magnetic flux
thrown onto the BH as the flux of open magnetic-field lines at the NS
polar caps \citep{GJ:1969, LyutikovMckinney:2011},
\begin{equation}
\phi = 2 \pi  B_{\rm NS} R^2_{\rm{NS}} \ \rm{sin}^{-1}\left(\frac{ R_{\rm{NS}} \Omega}{c}\right),
\end{equation}
where, in the single NS case, $\Omega$ is the NS spin angular
frequency, but here the light cylinder, and hence the footprint of
open field lines on the NS surface, is determined by the orbital
velocity in addition to the NS spin. Approximating $\Omega$ as the
orbital angular frequency near merger, Figure \ref{Fig:PostMerger}
plots the initial power available to the post-merger BH as a function
of BH spin.

Notice that the post-merger BZ power scales as $M^{-2}$ through
$R_H(S)$ whereas the usual BZ power scales as $M^2$. The BZ power
depends on the square of the magnetic flux deposited onto the BH,
which in the standard case, scales with the squared BH surface area $M^4$; 
adding also the dependence on horizon
angular velocity, which scales as $M^{-2}$, gives the usual $M^2$
scaling. In the BHNS merger case however, the magnetic flux is set not
by the BH size, but by the available flux brought in by the NS, so
indeed larger BHs emit less BZ power.

Such a post-merger event will likely generate a relativistically
beamed jet which peaks at maximum luminosity given by Figure
\ref{Fig:PostMerger} and then decays with the decaying BH
magnetosphere. If the BH can hold onto the magnetosphere for a long
enough time, such an event might generate a type of afterglow to the
BHNS merger.  Assuming that the post-merger signal begins at the same
time as fireball expansion, at merger, then the peak luminosity of the
post merger signal would be observed $R_{\rm{ph}}/c \sim 0.2
\ \rm{msec} \sqrt{B/10^{12}\rm{G}}$ before the blackbody fireball
emission. We mention this as it is of observational interest and an
avenue to pursue in developing the full portrait of the BH battery.

%%%%%%%%%%%%%%%%%%%%%%%%%%%%%%%%%%
%%%FIGURE Post Merger Lum
%%%%%%%%%%%%%%%%%%%%%%%%%%%%%%%%%%
%%%%%%%%%%%%%%%%%%%%%%%%%%%%%%%%%%%%%%%%%%%%%%%%

%%%%%%%%%%%%%%
%%Section: Observability
%%%%%%%%%%%%%%
\section{Observability}
\label{Observability}
The Fermi GBM \citep[GBM][]{FERMIGBM:2015} is well suited for detecting
the transients described above. It has an energy range of $0.008
\rightarrow 30$ MeV, capturing the peak of emission predicted for
binaries with $10^{12}$ to $\sim 10^{16}$ G NS magnetic-field
strengths (Eq. \ref{Eq:KTWein}). It has a $2 \mu$s timing resolution,
sufficient to resolve the $\gtrsim 1$msec bursts. The Fermi GBM also
operates with a nearly full-sky field of view (currently operating at
9.5 sr with a 10 sr goal), important for catching such possibly rare
transients.

We estimate the photon flux at the instrument by assuming emission
from a blackbody with Doppler boosted (Eq. \ref{Eq:Teff}) and
cosmologically redshifted temperature. The photon flux at the GBM is
\begin{align}
F_{\rm{obs}} &= 2 \pi \int^{\theta_c}_{0}\int^{\nu_{\rm{max}}}_{\nu_{\rm{min}}}{  \frac{2 \nu^2}{ c^2} \frac{\cos{\theta} \sin{\theta} d \nu d\theta}{\rm{exp\left[\frac{h \nu (1+z)}{ k T_{\rm{eff}}(\theta)}\right] - 1} }  }  \\ \nonumber
\theta_c &=\frac{R_{\rm{ph}}}{d_A(z)}  \\ \nonumber
d_A(z) &= \frac{c}{H_0} \int^z_0{\frac{dz'}{\sqrt{\Omega_M(1+z)^3 + \Omega_{\Lambda}}}}   \\ \nonumber
T_{\rm{eff}}(\theta)&= T_{\rm{ph}}\left[ \gamma \left( 1 - \frac{v}{c} \cos{\left( \frac{\pi}{2} \frac{\theta}{\theta_c}\right) } \right)\right]^{-1}
\end{align}
where $d_A$ is the angular diameter distance in the 2015 Planck
cosmology with $\Omega_M = 0.308$, $\Omega_{\Lambda} = 1-\Omega_m$,
and $H_0 = 67.8$ km s$^{-1}$ Mpc$^{-1}$
\citep{Planck:2015:CosmoParams}, and where integration is over the
solid angle of the photosphere at redshift $z$, and over the frequency
limits of the GBM. We use the minimum detectable flux for the GBM to
solve $F_{\rm{obs}}(z) = F_{\rm{min}}$ for the maximum observable
redshift to which BHNS transients could be observed. Using the GBM
on-board trigger sensitivity, $F_{\rm{min}} = 0.71$ cm$^{-2}$ s$^{-1}$
\citep{FERMIGBM:2015}, we find
\begin{align}
d^{\rm max}_M (B_{\rm NS} = 10^{12} \mbox{G}) &\sim 9 \  \mbox{Mpc};  \ \ \ \quad z^{\rm{max}} = 0.002   \nonumber \\ 
d^{\rm max}_M (B_{\rm NS} = 10^{13} \mbox{G}) &\sim 49 \  \mbox{Mpc};  \ \quad z^{\rm{max}} = 0.011   \nonumber \\ 
d^{\rm max}_M (B_{\rm NS} = 10^{14} \mbox{G}) &\sim 270 \  \mbox{Mpc};  \quad z^{\rm{max}} = 0.064   \nonumber \\
d^{\rm max}_M (B_{\rm NS} = 10^{15} \mbox{G}) &\sim 1.3 \  \mbox{Gpc};  \ \quad z^{\rm{max}} = 0.339   \nonumber \\
d^{\rm max}_M (B_{\rm NS} = 10^{16} \mbox{G}) & \sim 5.1 \  \mbox{Gpc};  \ \quad z^{\rm{max}} = 1.886,
\end{align}
which we have quoted in terms of the comoving radial distance $d_M$
and the corresponding redshift. The $\gtrsim 10^{13}$G binaries are
detectable out to beyond the initial LIGO volume, while only the
$\gtrsim 10^{14.5}$G binaries are detectable out to approximately the
advanced LIGO volume for BHNS mergers \citep[redshift
  $z\sim0.1$;][]{LIGORates:2010}.

To estimate the number of expected detections out to $z^{\rm{max}}$ we
need to know the rate of BHNS mergers as a function of $B_{\rm{NS}}$,
and we need to know what fraction of those mergers generate the signal
derived here. BHNS coalescence rates are computed by
Ref. \cite{LIGORates:2010}. They predict between $6 \times 10^{-4}$
and 1 BHNS coalescences per Mpc$^3$ per Myr with a most probable rate
of $0.03$ per Mpc$^3$ per Myr. Estimating the number of nondisrupting
BHNS mergers with a given NS magnetic-field strength is beyond the
scope of the present work. Instead, we parametrize the fraction of
BHNS mergers which generate the signal predicted here as $f_{\rm
  fb}(B_{\rm{NS}})$. Using the calculated maximum detection redshifts
we calculate the comoving detection volume. Using this maximum
detection volume, coalescence rates with $f_{\rm fb} = 1$, and a 10 sr
field of view, Table \ref{Tab:Rates} lists the expected number of
events that FERMI GBM could detect per year.
\begin{table}
\begin{center}
\begin{tabular}{ c | c | c | c }
    $B_{\rm{NS}}$  [G]       & Minimum              & Expected 			    & Maximum   \\
                   \hline 
 $10^{12}$  &   $1.4 \times 10^{-6}$      & $6.9 \times 10^{-5}$     & $2.3 \times 10^{-3}$  \\
 $10^{13}$  &   $ 2.4 \times 10^{-4}$     & $1.2 \times 10^{-2}$     & $0.4$  \\
$10^{14}$  &   $3.9 \times 10^{-2}$       & $2.0$			          & $66$	\\
$10^{15}$  &   $5.0$                              & $248$     		          & $8.3 \times 10^{3}$   \\
$10^{16}$  &   $267$                             & $1.3 \times 10^{4}$       & $4.5 \times 10^{5}$	 
 \end{tabular}
\caption{Expected number of Fermi GBM events in units of
  $\left[\mbox{yr}^{-1}\right]$ $f_{\rm fb}(B_{\rm{NS}})$ where
  $f_{\rm fb}(B_{\rm{NS}})$ is the fraction of BHNS coalescences with
  NS magnetic-field strength $B_{\rm{NS}}$ and which will not tidally
  disrupt the NS and will generate the signal predicted here. $B_{\rm{NS}}$
  is the NS surface magnetic-field strength.}
\label{Tab:Rates}
\end{center}
\end{table}

For BHNS binaries with $B_{\rm{NS}} \lesssim 10^{14}$G, these
optimistic, expected rates of detection drop below 1 per year. To
probe the binaries with $B_{\rm{NS}} \gtrsim 10^{13}$G at a rate of
$\sim 1.0 f_{\rm{fb}}$ yr$^{-1}$, future x-ray instruments must have
full-sky sensitivities of $\sim10\times$ the FERMI GBM. They must have
sensitivities $\sim600\times$ the GBM to reach $B_{\rm{NS}} \gtrsim
10^{12}$G binaries at the same rate.

Assuming our model roughly captures the BHNS luminosity and spectrum, there are two options for BHNS mergers with $B_{\rm{NS}} \gtrsim 10^{14}$G. Either we have already observed the high-magnetic-field BHNS fireballs as a subclass of short gamma-ray bursts (sGRBs), or we have not, and the fraction of nondisrupting BHNS binaries with such magnetic fields $f_{\rm{fb}}$ is very small.

The BHNS fireball could compose a subclass of the sGRB population if
a, yet unknown, mechanism saturates NS field strengths to maximal
$\geq 10^{15}$G values near merger, then the rates predicted here
become comparable to the inferred (beaming angle dependent) rates of
sGRBs, $8\rightarrow 1100$ Gpc$^{-3}$ yr$^{-1}$ from Swift measurements
\citep{SWIFTsGRBrates:2012}. The analysis of \S \ref{Fireball}
allows emission from $\sim10^{15}$G fireballs to be of order seconds,
consistent with sGRB time scales.

Alternatively, evidence has been found that a class of sGRBs, making
up 10 to 25 percent of the total, may be at a near $z \leq 0.025$
\cite{Tanvir:2005}. These would be a different class than those sGRBs
for which distances can be measured out to a Gpc through afterglows
\citep[{\em e.g.},][]{Berger:2005}. The implication is that a class of
sGRBs has a much lower luminosity engine, which could be powered by
the $B_{\rm{NS}} \sim 10^{13}$G BHNS transients discussed
here. This possibility, however, requires an
explanation for increased rates of BHNS mergers in the local universe.

If the BHNS fireball is not a subset of the observed GRB population,
then, based on the present nondetection, we may place limits on the
fraction of binaries which carry $B_{\rm{NS}} \gtrsim 10^{14}$G, to
merger. Using the expected rates and the total operation time of the
GBM at its current sensitivity ($\sim5$ years) we find that
$f_{\rm{fb}}(\geq 10^{15} \rm{G}) \lesssim 10^{-3}$ and
$f_{\rm{fb}}(\geq 10^{16} \rm{G}) \lesssim 10^{-4}$. Where the
inequalities assume that $f_{\rm{fb}}$ is a steeply decreasing
function of magnetic-field strength for $B_{\rm{NS}} > 10^{14}$G.

Another possibility is that these upper limits for the luminosity of
the signal are indeed overestimates and mechanisms such as screening
in the magnetosphere greatly damp power output; continued electromagnetic, as well as future gravitational wave, observations
will test this. Concurrently, further modeling of the BHNS
magnetosphere would hone the expected signal and the derived rates of
detection.

The above analysis relies on a choice of $R_0=GM/c^2$ for the size
scale of energy injection. This is a natural choice, however we
discuss briefly the dependence of our results on
injection radius. If we go with a large value of $R_0=2GM/c^2$, then less energy is injected
over a larger volume and the initial temperature of the fireball drops
to $18$ keV $(B_{\rm NS}/10^{12} G )^{1/2}$ from our fiducial $85$ keV
$(B_{\rm NS}/10^{12} G )^{1/2}$ for $R_0=GM/c^2$. This corresponds to
a peak black body temperature of $52$ keV $(B_{\rm NS}/10^{12} G
)^{1/2}$, down from the fiducial $0.24$ MeV $(B_{\rm NS}/10^{12} G
)^{1/2}$. These lower energies are still within the energy range of
the Fermi GBM, but a combination of less injected energy, smaller
photosphere sizes (Figure \ref{Fig:Rphot}) (and hence smaller
expansion speed at the photosphere) decrease the maximum observable
distance of the fireball by a factor of $\sim3$ and also decreases the
expected rates (Table \ref{Tab:Rates}) by one to two orders of
magnitude.

\section{Conclusion}
We have used BH-battery energetics to argue that near merger, a BHNS
will produce an electromagnetic transient.  A spectrum of high-energy
($\sim$ TeV) curvature radiation will escape the magnetosphere before
the last $0.1$s $(B/10^{12}\rm{G})$ of inspiral. This signature will
only reach luminosities of $\sim 10^{38} \rm{erg} \ \rm{s}^{-1}
\left(B/10^{12}\rm{G}\right)^{1/2}$ before being quenched by
pair production and fueling the more luminous fireball transient.  The
expanding fireball will become transparent and emit as a blackbody in
the x-ray to $\gamma$-ray range for of order $10^{-3} \rightarrow 10$
seconds depending on the NS magnetic-field strength. The observed
luminosity can peak at $10^{45}$ erg s$^{-1}$ for a $10^{12}$G NS
magnetic field or up to $10^{53}$ erg s$^{-1}$ for magnetar strength
fields. If the BH can hold onto the NS magnetic fields after merger
through a slow decay of the magnetosphere
\citep{LyutikovMckinney:2011}, a spinning remnant BH could power a
relativistic jet with bolometric luminosity up to 2 orders of
magnitude lower than the fireball luminosity, peaking at $ \sim 0.2
\ \rm{msec} \sqrt{B/10^{12}\rm{G}}$ before the observed fireball
emission, and decaying on the unknown resistive timescale of the
magnetosphere.

The prospects for detecting the bright, fireball transient are
dependent on the (unknown) distribution of NS magnetic-field strengths
$B_{\rm{NS}}$ at merger. To explore these prospects, we have left the NS surface 
magnetic-field strength as a free parameter. Conversely, BHNS merger rates allow our model to put constraints on $B_{\rm{NS}}$ at merger. Given predicted BHNS merger rates, the majority of BHNS mergers must have $B_{\rm{NS}} > 10^{14}$G to be
detectable by Fermi GBM at the rate of $\sim 1$ yr$^{-1}$. If
$B_{\rm{NS}}\lesssim 10^{12}$ at merger, as might be expected from the
observed pulsar magnetic-field strengths \citep{PulsarsRev:1991}, a
future x-ray instrument would need a full-sky sensitivity of $\gtrsim
600$ the present FERMI GBM capabilities to detect these EM signatures
of BHNS coalescence. If ordered magnetic fields are amplified to
$\gtrsim 10^{15}$G at merger, then expected FERMI GBM detection rates
for the signature in this study climb to rival the gamma-ray burst
rate, and may be a subclass of sGRBs \citep{SWIFTsGRBrates:2012}.

Any observation of a BH-battery transient would be exciting in its own
right.  With advanced LIGO now operational, the EM counterpart to BHNS
coalescence has additional payout potential, offering unique
information to extend the astronomical reach of the gravitational-wave
observatories.

\acknowledgements The authors thank Andrei Beloborodov, Brian Metzger,
and Sean McWilliams for useful discussions. The authors also thank the anonymous referee 
for comments that improved the manuscript.  D.J.D. acknowledges support
from a National Science Foundation Graduate Research Fellowship under
Grant No. DGE1144155.  J.L. thanks the Tow Foundation for their
support. J.L. was also supported by a Guggenheim Fellowship and is a
Chancellor's Fellow at Chapman University. This research was undertaken, in part, thanks to funding from the Canada Research Chairs program. NWM was supported in part by the Natural Sciences and Engineering Council of Canada.

\appendix
\section{Parameter Dependence of Curvature Spectra}
\label{AppendixA}
\begin{widetext}

 %%%%%%%%%%%%%%%%%%%%%%%%%%%%%%%%%%
%%%FIGURE Appendix Curv and Synch Spectra 
%%%%%%%%%%%%%%%%%%%%%%%%%%%%%%%%%%
\begin{figure}
\begin{center}$
\begin{array}{c c c}
\hspace{-10pt}
\includegraphics[scale=0.24]{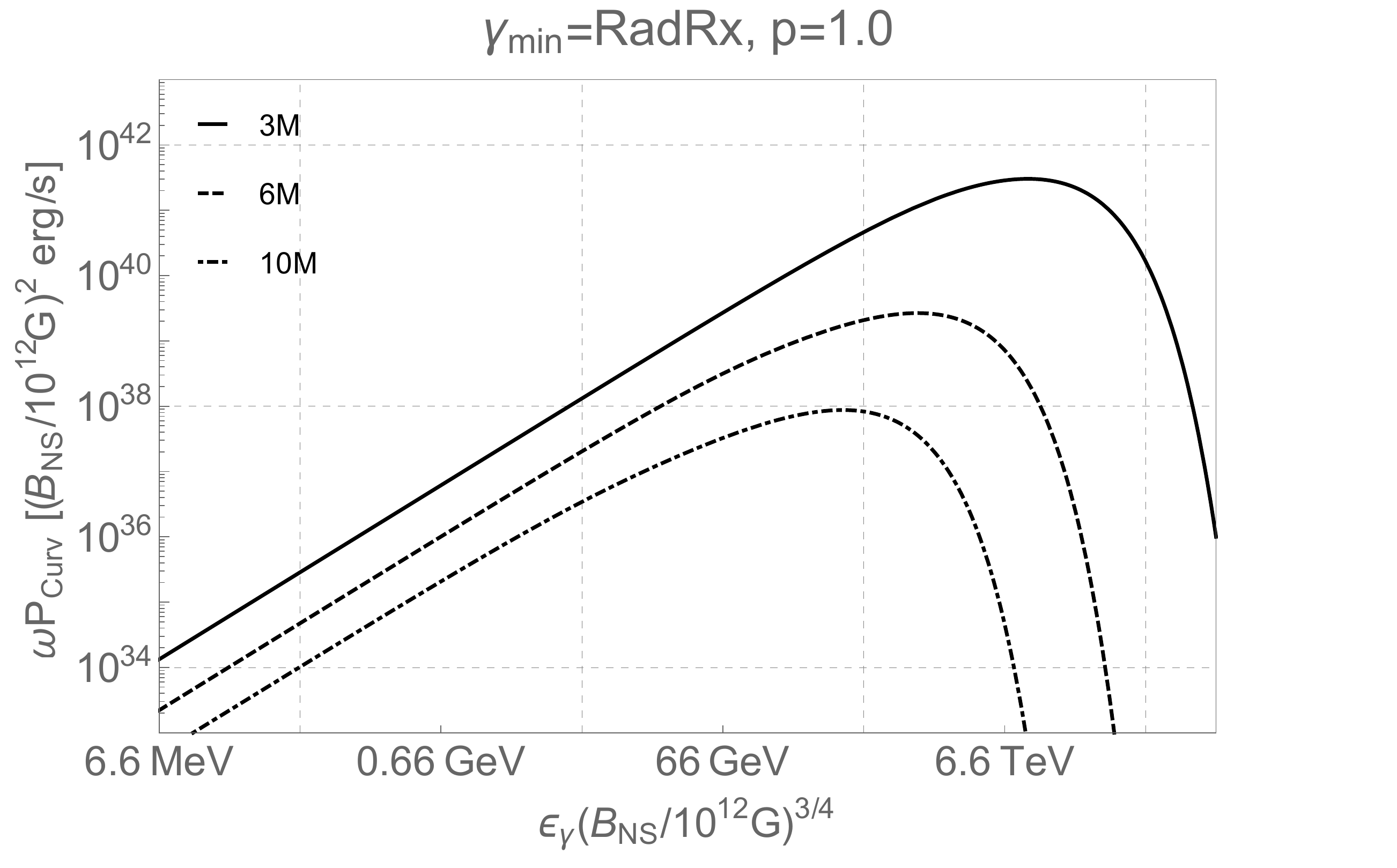} &
\hspace{-10pt}
\includegraphics[scale=0.24]{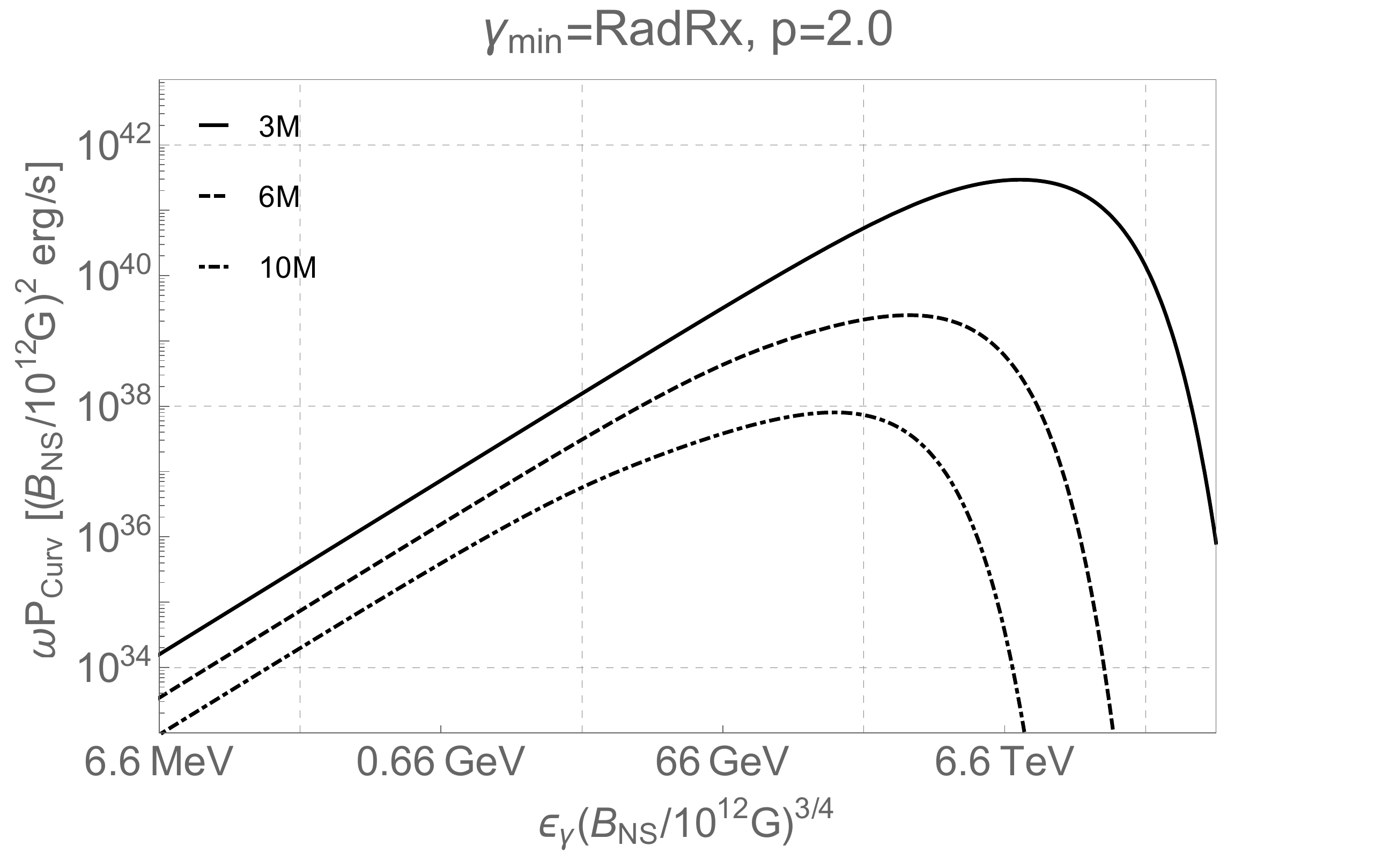} &
\hspace{-10pt}
\includegraphics[scale=0.24]{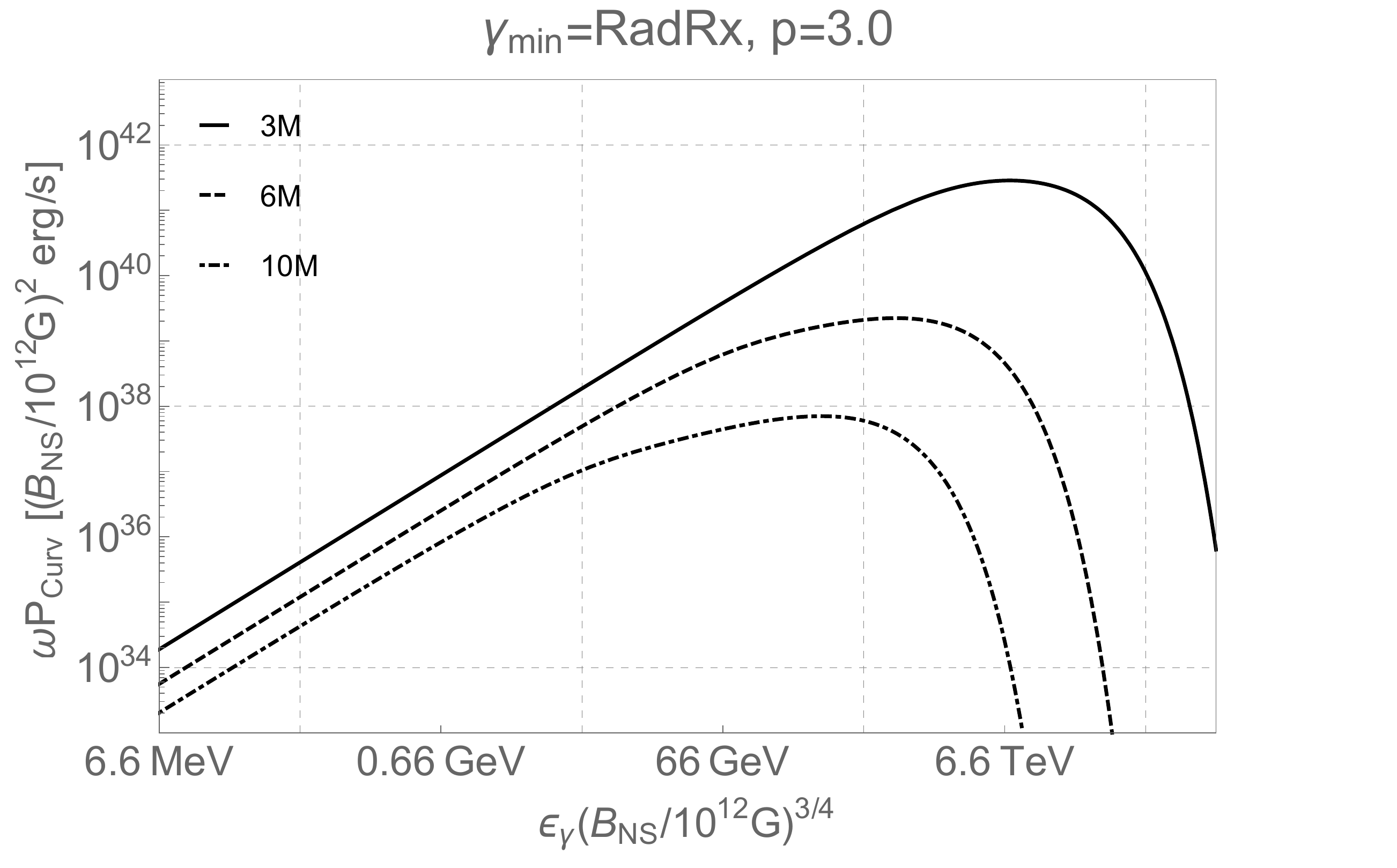}  \\
\hspace{-10pt}
\includegraphics[scale=0.24]{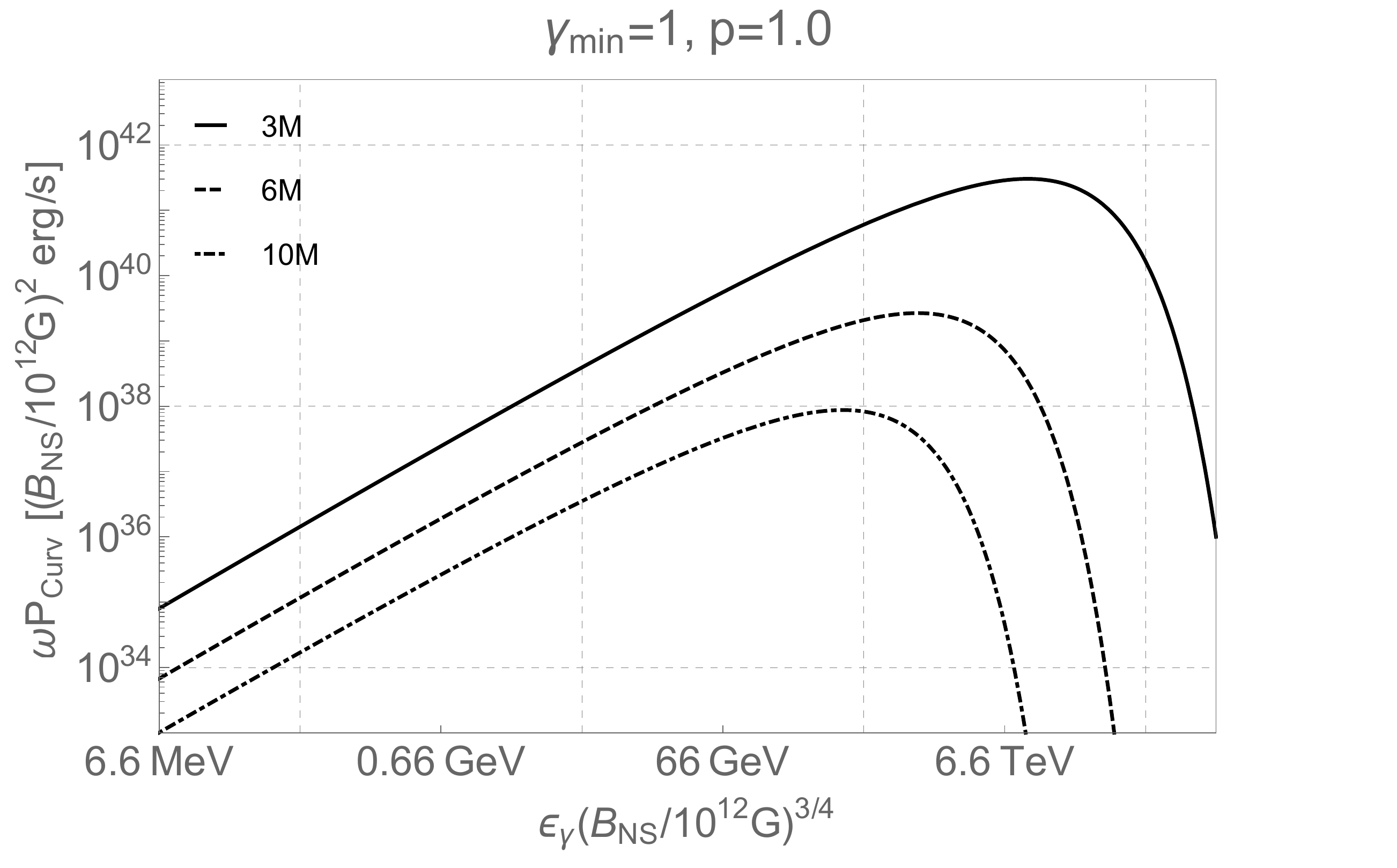} &
\hspace{-10pt}
\includegraphics[scale=0.24]{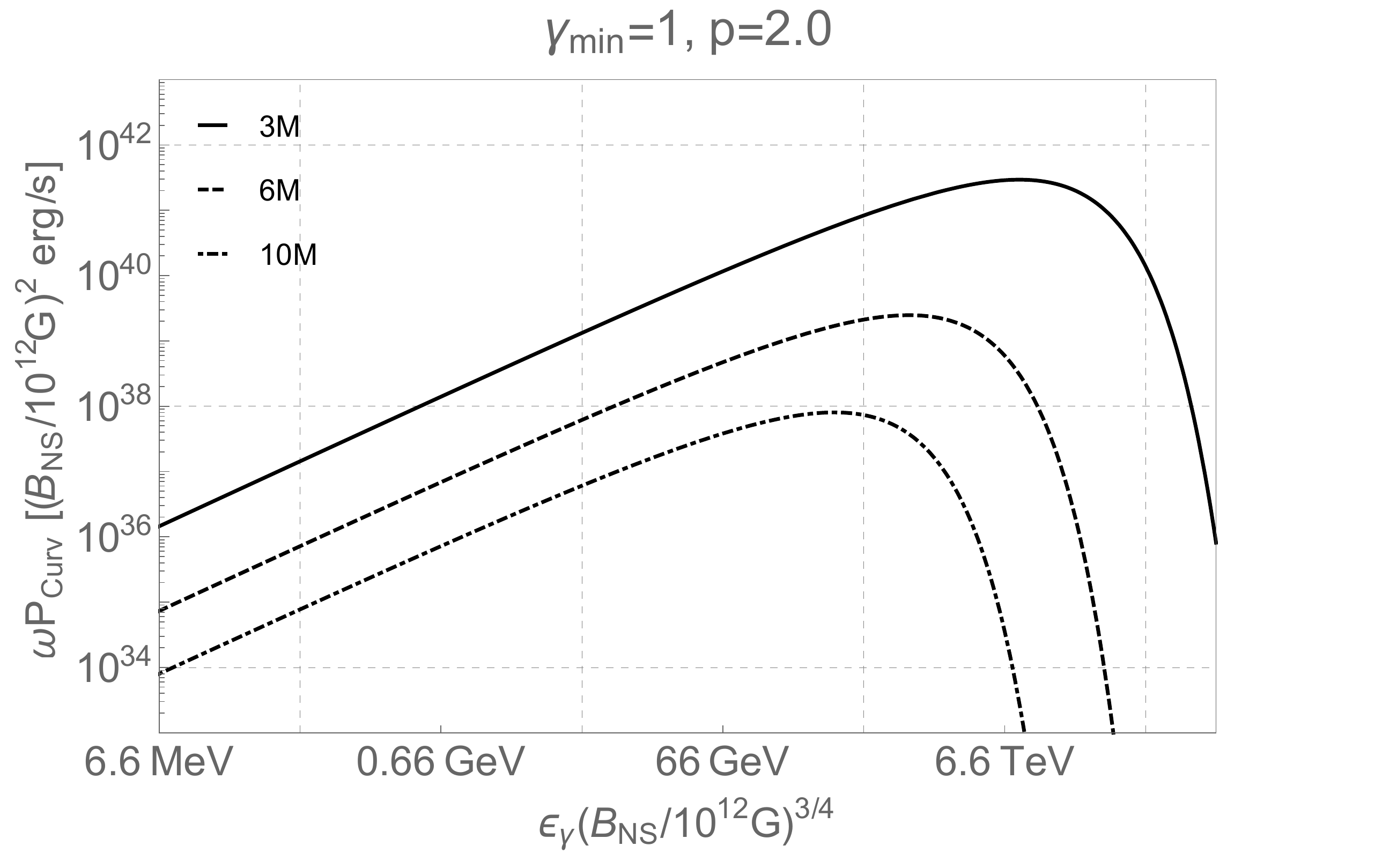} &
\hspace{-10pt}
\includegraphics[scale=0.24]{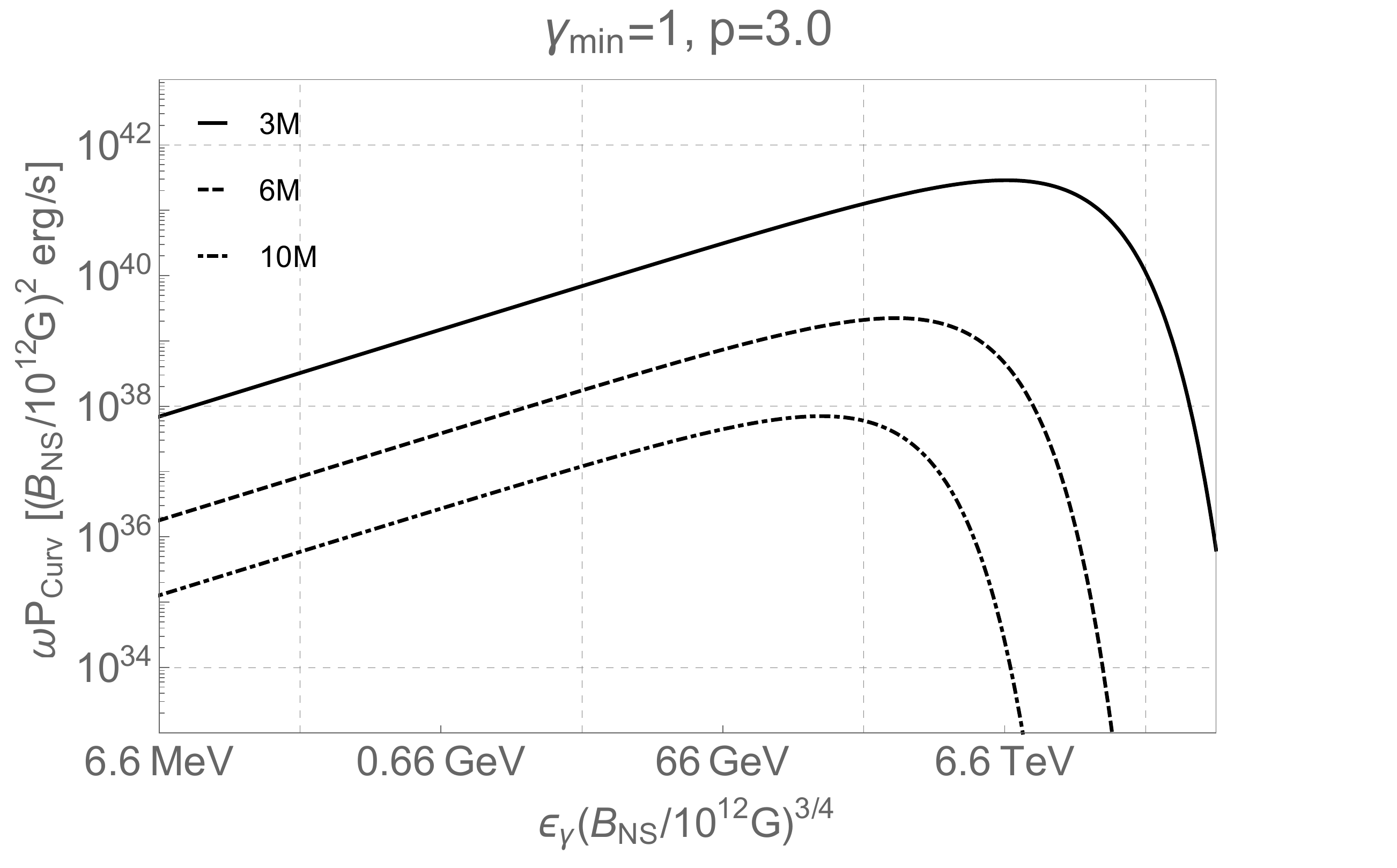} 
\end{array}$
\end{center}
\caption{The spectrum of primary curvature radiation at times
  corresponding to binary separations of $10M$, $6M$, and $3M$
  (dot-dashed, dashed, solid). Each panel is for the labeled minimum
  electron Lorentz factor and power law index $p$ of electron
  energies. $\gamma_{\rm min} = $RadRx refers to the radiation
  reaction limited Lorentz factor at the point of weakest electric
  field in the region connecting NS and BH (of order a few to 10 times
  smaller than the maximum Lorentz factor near merger).}
\label{Fig:AppCurvSpectra}
\end{figure}
%%%%%%%%%%%%%%%%%%%%%%%%%%%%%%%%%%%%%%%%%%%%%%%%

\end{widetext}

Figure \ref{Fig:AppCurvSpectra} plots the curvature radiation spectra,
identical to Figure \ref{Fig:CurvSpectra_p2}, but for different values
of the electron-energy power law index $p$, and the minimum electron
Lorentz factor in the magnetosphere, $\gamma_{\rm min}$. We vary $p$
from $1.0$ to $3.0$. We choose minimum Lorentz factors which bracket
the range of plausible values: $\gamma_{\rm min}=1$, and a minimum
radiation-reaction limited Lorentz factor which we compute with
Eq. (\ref{Eq:GamMax}) but with electric field at the edge of the
binary orbital light cylinder ($\Omega_{\rm orb}/c$) that falls off
from its horizon value as $r^{-2}$ \citep{MPBook}. Near merger this is
only a few times smaller than the maximum $\gamma$ computed form the
horizon electric fields.
 
]

\bibliography{NSBH}
\end{document}